%% file: bubble_scattering02.tex
\documentclass{osa-article}

\journal{osajournal}

\articletype{Research Article}

\usepackage{bm} %
\usepackage{amsmath}
\usepackage{mathrsfs}
\usepackage{siunitx}
\usepackage{import}

\usepackage{ulem}

\def\ie{\textit{i.e.}\ }  %
\def\eg{\textit{e.g.}\ }  %
\def\etal{\textit{et al.}\ }  %
\def\e{\text{e}}
\def\graphpath{./}
\graphicspath{\graphpath}

\def\dixmoins#1{10\textsuperscript{-#1}}

\begin{document}

\title{Modeling multiple scattering transient of an ultrashort laser pulse by spherical particles}

\author{Geoffroy Chaussonnet\authormark{1}, Lo{\"i}c Mees\authormark{2}, Milo\v{s} \v{S}ormaz\authormark{3}, Patrick Jenny\authormark{3}, Philippe M. Bardet\authormark{1}}

\address{\textsuperscript{1}George Washington University, 2121 I St NW, Washington, DC 20052, USA}
\address{\textsuperscript{2}CNRS, Univ Lyon, Ecole Centrale de Lyon, INSA Lyon, Univ Claude Bernard Lyon 1,LMFA, UMR5509, 69134, Ecully, France}
\address{\textsuperscript{3}Institute of Fluid Dynamics, ETH Zurich, Sonneggstra\ss e 3, 8092 Z{\"u}rich, Switzerland}

\email{\authormark{*}gchaussonnet@gwu.edu} %

\begin{abstract}
The multiple scattering of an ultrashort laser pulse by a turbid dispersive medium (namely a cloud of bubbles in water) is investigated by means of Monte Carlo simulations. The theory of Gouesbet and Gr{\'e}han [Part. Part. Syst. Charact. {\bfseries 17} 213-224 (2000)] is used to derive an energetic model of the scattering transient.
It is shown that the spreading and extinction of the pulse can be decoupled from the transient of scattering,
which allows to describe each phenomenon individually.
The transient of scattering is modeled with the Lorenz-Mie Theory and thus is also valid for a relative refractive index lower than one, contrary to the Debye series expansion which does not converge close to the critical angle.
This is made possible after the introduction of a new physical object, the Scattering Impulse Response Function (SIRF) which allows to detect the different modes of scattering transient, in time and direction.
The present approach is more generic, as it enables to simulate clouds of air bubbles in water, which was not possible previously.
Two different approaches are proposed within the Monte Carlo framework. The first is a pure Monte Carlo approach where the delay due to the scattering is randomly drawn at each event, while the second is based on the transport of the whole scattering signal. 
They are both embedded in the Monte Carlo code Scatter3D [JOSA A {\bfseries 24}, 2206-2219 (2007)].
Both models produce equivalent trends and are validated against published numerical results.  They are then applied to the multiple scattering of ultra short pulse by a cloud of bubble in water in the forward direction.
The pulse spread due to the propagation in water is computed for a wide range of traveled distances and pulse durations, and the optimal pulse duration is given to minimize the pulse spread at a given distance.
The main result is that the scattered photons exit the turbid medium earlier than the ballistic photons and produce a double peak related to the refraction in the bubble. This demonstrates the possibility to develop new diagnostics to characterize dynamic bubbly flows.
\end{abstract}

\section{Introduction}

The characterization of turbid media appears in a very broad spectrum of applications like detection of cancerous cells in organic tissues \cite{kunnen_application_2015}, characterization of dense fuel spray in combustion engines \cite{linne_ballistic_2009}, or atmospheric optics \cite{ishimaru_wave_1978} to name a few.
Such media are by definition optically thick, which means that when light beams cross turbid media, most of the photons undergo multiple scattering and very few of them are unaffected. This property makes turbid media look opaque and conventional optical diagnostics to characterize dynamic media such as shadowgraphy, laser diffraction or laser Doppler anemometry perform poorly \cite{linne_imaging_2013}. Solutions were developed for stationary media based, for example, on periodic polarization modulation \cite{mujumdar_imaging_2004}, or structured illumination and Fourier filtering \cite{berrocal_high-contrast_2016}.

The development of ultrashort laser pulse below the picosecond in combination with ultrafast time gating offers an alternative approach, commonly called ballistic imaging \cite{galland_time-resolved_1995}. The method consists in recording the time of arrival of an ultrafast laser pulse through the medium; it has been demonstrated in gas flows ladden with droplets.  In this configuration and with adequate time resolution, the signal reveals a primary peak made of ballistic (no interaction with the suspension) and snake (only diffraction) photons, followed by a second smoother peak made of scattered photons. The resolution of these peaks allows to identify the global characteristics of the cloud of scatterers such as the concentration, the mean diameter, or the width of the size distribution \cite{calba_monte_2006}.
This technique was experimentally demonstrated 
by Calba \etal \cite{calba_ultrashort_2008} with polystyrene particles immersed in water
and by Linne \etal \cite{linne_ballistic_2009} in the case of a dense fuel spray.

While not studied to date, ballistic imaging could be of interest for dense bubbly flows, particularly where it is important to record the smallest bubbles.  This includes a broad range of fields, such as environmental flows, nuclear thermal hydraulics, hydraulics, and naval hydrodynamics. This configuration differs from the ones mentioned above in two aspects.
First, the scatterers (air bubble) have a refractive index lower than that of the propagation medium (water), which means that refracted photons are faster than ballistic ones \cite{chaussonnet_scattering_2020}. As a result it can be expected that the peaks of the transmitted light signal might appear in a different order than those with water droplet in air.
Second, due to the significant variations of the water refractive index in the visible light, dispersion and extinction of the pulse must be accounted for.
The present paper proposes to model the transient multiple scattering of an ultrashort laser pulse with the LMT to investigate the effect of bubbles characteristics on the output temporal signal.
An important consequence of the relative refractive index lower than one (air bubbles in water) is the presence of the critical angle of refraction where the Debye series do not converge.
In the pioneering work of Calba \etal \cite{calba_monte_2006} the scattering transient was described by the Debye expansion of Lorenz-Mie series, which allows to separate each order of refraction, to facilitate their identification in time and direction, and finally to greatly simplify the modeling.
Since the Debye series would not converge in the present configuration, it is necessary to modify the modeling strategy of the scattering transient.
In the present work we use the full Lorenz-Mie Theory (LMT) to have an exact description of the scattering at any angle. In order to differentiate the peaks of energy in time and direction, we introduce a new physical object, the Scattering Impulse Response Function (SIRF) which is the virtual response of the scatterer to an infinitely short pulse. This new approach leads to a better time-separation of the scattering modes, and enables us to build a more generic model. This is one of the novelties of this work.
\\
Multiple scattering in a turbid medium can be modeled by two approaches \cite{ishimaru_wave_1978}.
The first one starts from basic differential equations (Maxwell or the wave equation), then incorporates the scattering and absorption properties of particles. It leads to differential or integral equations describing statistics such as variances or correlation functions, which account for all phenomena such as multiple scattering, diffraction, and interferences.
This approach is mathematically rigorous, but the resolution of the equations is in practice computationally prohibitive.
Alternatively, the radiative transfer theory is based on the transport of energy through a medium containing particles. It is described by the Radiative Transfer Equation (RTE) which is equivalent to Boltzmann's equation from the kinetic theory of gases.
It neglects self interactions of the electromagnetic field during propagation, such as interference, enabling the superposition principle for intensities and powers.
The eletromagnetic effects due to the particles such as diffraction and interference at the particle scale are accounted by an appropriate modeling of the scattering, such as the LMT.
This approach is thus phenomenological and allows to solve various practical problems such as atmospheric and underwater visibility, marine biology, and photographic
emulsions \cite{ishimaru_wave_1978}.
It was shown recently that when averaged over a sufficiently long period of time, Maxwell's electromagnetic theory for multiple wave scattering in discrete random media can be related to the radiative transport theory \cite{mishchenko_vector_2002,mishchenko_microphysical_2003}.

The radiative transfer theory was selected for the present study.
To accurately predict the propagation of an ultrashort pulse through a turbid medium seeded with large scatterers (1 - \SI{100}{\micro\meter}), the transient of scattering must be accounted for \cite{roze_interaction_2003}, which is the case here.
Therefore we need to incorporate the transient of the LMT into the scattering model of the RTE.
To do this we use the theory of Gouesbet and Gr{\'e}han \cite{gouesbet_generic_2000} that describes far field temporal single scattering of an electromagnetic (EM) pulse.
The structure of this study is as follows.
In Section~\ref{sec_methodo} we derive the multiple scattering transient in a dispersive medium based on the theory of Gouesbet and Gr{\'e}han \cite{gouesbet_generic_2000}.
The results are then adapted to the radiative transport theory and the Monte Carlo method is presented in Section~\ref{sec_montecarlo}.
The spread of an ultrashort pulse in water is discussed in Section~\ref{sec_pulse_deformation}.
Numerical aspects of the model are investigated in Section~\ref{sec_SSIRF}, and the present approach is validated in the case of single scattering in Section~\ref{sec_valide_energy_approx}.
Finally, the transient of multiple scattering is compared to other numerical simulations from the literature in Section~\ref{sec_transient_multi_energy} followed by more realistic simulations in Section~\ref{sec_num_exp}.

\section{Transient scattering due to an electromagnetic pulse \label{sec_methodo}}

This section details the mathematical aspect of the transient of multiple scattering in a dispersive medium and how it can be decomposed into convolution products. First the spreading of a pulse is presented, followed by single scattering and multiple scattering.

\subsection{Pulse spreading in a dispersive medium}
\begin{figure}[!htb]
	\begin{center}
	\def \svgwidth {0.7\textwidth}
	{\scriptsize
	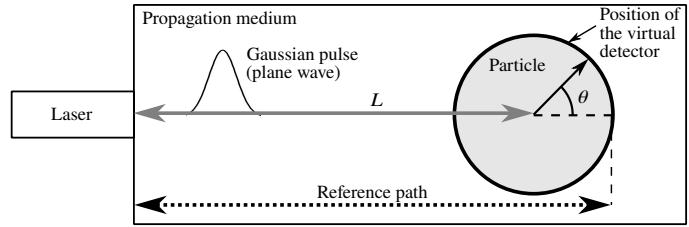
	}
	\caption{Sketch of the modeled configuration.}
	\label{fig_sketch_geom}
	\end{center}
\end{figure}

Let $\bm{\psi}$ be the incident electric (E) or magnetic (H) field for a pulsed plane wave in a dispersive medium. 
We assume that a laser located at $z=-L$ (Fig.~\ref{fig_sketch_geom}) produces a Gaussian pulse whose peak is at $z=-L$ at the time $t = -T \equiv -L/v_g$ where $v_g$ is the group velocity defined by $v_g = {\partial \omega}/{\partial k}$ at the carrier frequency $\omega_0$.

The temporal envelop $g$ of the pulse at the laser location and its Fourier Transform (FT) $G$ read:
\begin{equation}
g(t,z=-L) = \exp \left[ - \left( \frac{t + T} { \gamma} \right)^2 \right]
\quad \text{and} \quad
G(\omega) = \frac{\gamma}{\sqrt 2} \text{e}^{-(\gamma \omega/2)^2}
\label{eq_pulse_FT_pulse}
\end{equation}
where $\gamma$ is a time constant related to the Full Width at Half Maximum (FWHM) $\Delta t$ for a Gaussian envelope:
\begin{equation}
\gamma = \frac{\Delta t} {2 \sqrt{\log(2)}}
\label{eq_gamma}
\end{equation}
We assume that the center of a particle is located at $z=0$. With the definition of $L$, the peak of the pulse would reach $z=0$ at $t=0$ if no particle were here.
The incident plane wave at the particle location can be expressed in the time domain by the Inverse Fourier Transform (IFT):
\begin{equation}
\bm{\psi}(t,z) = \frac{\bm{\psi}_0}{\sqrt{2 \pi}}  \int_{0}^{+\infty} \text{e}^{i (\omega t - k z )} G(\omega-\omega_0) \,  \text{e}^{i (\omega T - k L)}  \, \mathrm d \omega
\label{eq_incident_wave_disp_new_frame}
\end{equation}
where $k=\omega/c$ is the wave vector and $\bm{\psi}_0$ is the amplitude of the incident wave.
The argument of the second complex exponential in Eq.~\ref{eq_incident_wave_disp_new_frame} can be written as $i L(\omega/v_g - k)$ which represents of the pulse spreading over the distance $L$ between the laser source and the particle.
It is convenient to gather the pulse and its propagation to express the FT of the chirped pulse as:
\begin{equation}
G'(\omega-\omega_0,L) =  G(\omega-\omega_0) \,  \text{e}^{i L (\omega / v_g- k)} 
\label{eq_deformed_pulse_FT}
\end{equation}
Note that in a dispersive and absorbing medium the wave vector $k$ is complex and has a non-linear dependency on the frequency $\omega$.
The envelop $g'$ of the chirped pulse is the slowly varying signal from Eq.~\ref{eq_incident_wave_disp_new_frame}. Its expression is given by taking out the carrier oscillations ($\omega_0, \Re(k_0)$) from Eq.~\ref{eq_incident_wave_disp_new_frame}.
In the following, we will consider the envelop $g'$ of the chirped pulse after an optical path of length $L$ in the dispersive medium:
\begin{equation}
g'(t,L) = \frac{1}{\sqrt{2 \pi}}  \int_{0}^{+\infty} \text{e}^{i [(\omega - \omega_0)(t-L/v_g) - (k - \Re(k_0))L]} G(\omega-\omega_0)  \, \mathrm d \omega
\label{eq_envelop_disp_new_frame}
\end{equation}
where $\Re(k_0)$ is the real part of $k_0$. 

\subsection{Transient of single scattering}

\begin{figure}[!htb]
	\centering
	\includegraphics[width=\columnwidth,keepaspectratio]{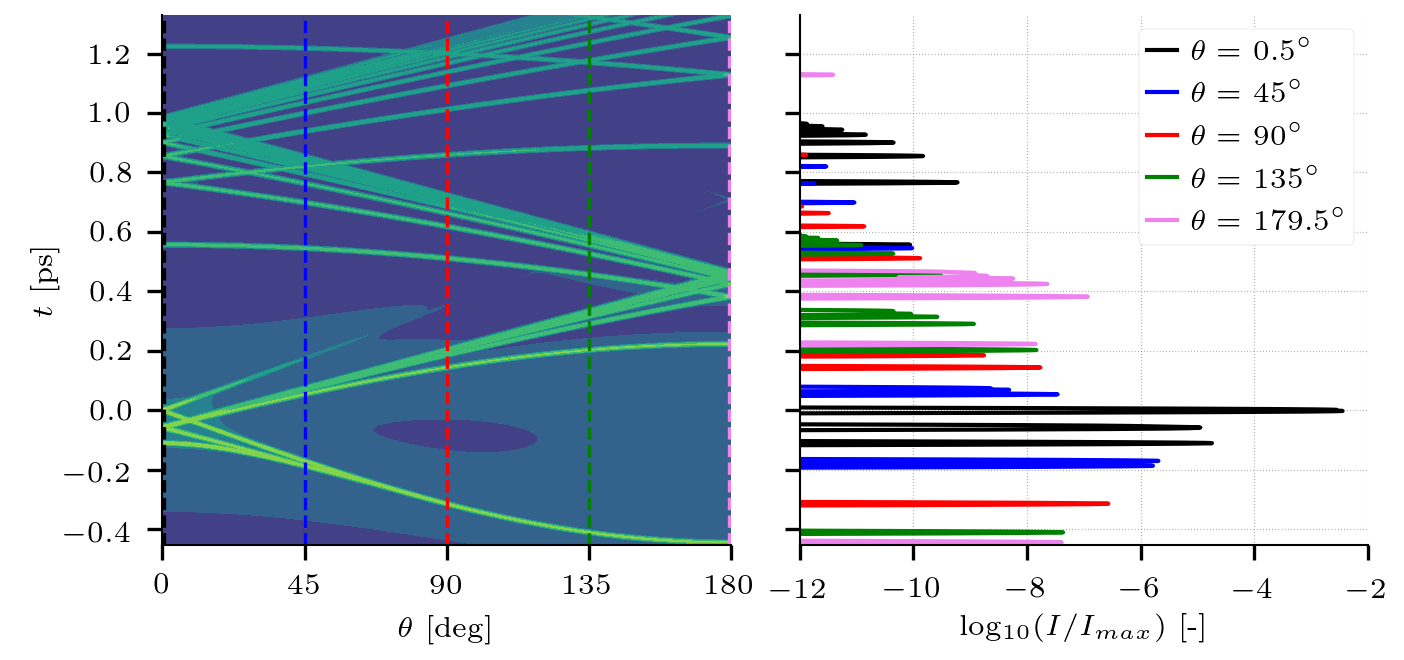}
	\caption{Left: ($\theta,t$)-map of the SSIRF for a bubble of \SI{100}{\micro\meter} ($x \approx 500$) and $\Delta t_v=4\pi/\omega_0$. Right: corresponding time signal at different angles.}
	\label{fig_illustre_SSIRF_time_signal}
\end{figure}

According to the theory of Gouesbet and Grehan \cite{gouesbet_generic_2000}, the transient of the scattering is expressed in the frequency domain and transformed back in the time domain by an IFT:
\begin{equation}
\bm{\psi}^s(t,\theta,L) = \mathscr{F}^{-1} [ G'(\omega-\omega_0,L)  \, \bm{\psi}^{cw}(\omega,\theta) ]
\label{eq_scatter_wave_disp_start}
\end{equation}
where $\mathscr{F}^{-1}$ represent the IFT and the superscript $cw$ stands for a continuous wave.
The term $\bm{\psi}^{cw}(\omega,\theta)$ represents the scattered field in the polar direction $\theta$ for a continuous monochromatic illumination at the frequency $\omega$, and it is determined from the LMT.
Within the LMT, $\bm{\psi}^{cw}$ depends only on the size parameter and the relative refractive index, respectively:
\begin{equation}
x \equiv {\pi \, d_s \, n_{\text{pm}}} / {\lambda_0} = r_s \, n_{\text{pm}} \,  \omega / c_0
\quad \text{and} \quad
m \equiv {n_{\text{s}}} / {n_{\text{pm}}}
\end{equation}
where $d_s$ and $r_s$ are the scatterer diameter and radius, and $\lambda$ the wavelength in vacuum of the incident wave. The terms $n_{\text{pm}}$ and $n_{\text{s}}$ are the complex refractive indices of the propagation medium and the scatterer, respectively.
The size parameter $x$ depends on $\omega$, and so does $m$ for a dispersive medium, which means that $\bm{\psi}^{cw}$ implicitly depends on $\omega$.
In the following we drop the dependency of $\bm{\psi}^{cw}$ on $x$ and $m$ for the sake of clarity, but we keep the implicit dependency on $\omega$.\\
The term $\bm{\psi}^s(t,\theta,L)$ in Eq.~\ref{eq_scatter_wave_disp_start} is the far field EM wave scattered by the particle in the polar direction $\theta$ at time $t$. The parameter $L$ mentions that before impacting the scatterer, the pulse traveled a distance $L$ in the dispersive medium. In the present expression we suppose that the virtual detector is located on the scatterer surface so that the pulse spreading after the scattering is not considered. 
In \cite{mees_time-resolved_2001}, the time $t$ is chosen according to a time of reference given by the path traveling in propagation medium only from the laser to the location of the center of the particle, then to a virtual detector located in the far-field.
However we define here the reference time as the propagation time from the laser source to the further boundary of the scatterer (Fig.~\ref{fig_sketch_geom}).
This can be approximately regarded as if a virtual detector were located at the surface of the particle and were recording the far-field scattering signal for all directions $\theta$ (Fig.~\ref{fig_sketch_geom}). Hence, $t=0$ when the maximum of the pulse reaches the position of the scatterer further boundary.
Also, contrary to the study by \cite{mees_time-resolved_2001}, we choose the sign of the time so that it increases as it elapses. Thus we define $t'$ as:
\begin{equation}
t' = \frac{d_s}{v_g} - t
\label{eq_time_shifted}
\end{equation}
In this case, $t'<0$ represents an optical path shorter than the reference path. 
In the following, we drop the prime symbols ($'$) from $t'$ for the sake of clarity and we write $t\equiv t'$.
Finally, the scattered field is given by:
\begin{equation}
\bm{\psi}^s(t,\theta,L) = \frac{1}{\sqrt{2 \pi}} \int_{\omega_{\min}}^{\omega_{\max}} G(\omega-\omega_0)  \text{e}^{i (\omega T - k L)} \, \bm{\psi}^{cw}(\omega,\theta) \, \text{e}^{i \omega t} \, \mathrm d \omega
\label{eq_scatter_wave_disp}
\end{equation}
where $\omega_{\min}$ and $\omega_{\max}$ are the bounds where $G(\omega)$ is significantly larger than 0, and $T=L/v_g$.
\\\\
Note that when $\omega$ is negative we use the relation $\bm{\psi}^{cw}(\omega)  = \bm{\psi}^{cw*}(-\omega)$, where the superscript~$*$ stands for the complex conjugate.
In the following, the scattered intensity is expressed as $I_i = |\bm{\psi}^s(t)|^2$ where the index $i$ can take the symbols $1$ and $2$ for the electric field in the incident and in the perpendicular to the planes, respectively. When no index is mentioned, the EM wave is assumed non-polarized and $I=(I_1+I_2)/2$.
\\Now we inspect each terms of Eq.~\ref{eq_scatter_wave_disp} in the frequency domain, where it is a product of three terms:
\begin{equation}
\mathcal{F} [\bm{\psi}^s](\omega,\theta,L) =
\underbrace{G(\omega-\omega_0)} _{\text{Pulse}}
\ 
\underbrace{\text{e}^{i (\omega T - k L)}}_{\text{Dispersion}}
\ 
\underbrace{\bm{\psi}^{cw}(\omega,\theta)}_{\text{Scattering}}
\label{eq_scatter_wave_disp_TF}
\end{equation}
The terms of the RHS are, in order of appearance, (i) the FT of the pulse as it exits the laser, (ii) the FT of the dispersion operator, which describes the spreading of the pulse from the laser to the particle, and (iii) the scattering operator.
We can make several remarks on Eq.~\ref{eq_scatter_wave_disp_TF}.
First, when the medium is non-dispersive, the dispersion operator is reduced to $\mathscr{D}(\omega) = 1$.
Second, contrary to the pulse and the dispersion terms, the scattering term $\bm{\psi}^{cw}$ is not the result of a FT, but was derived by Mie \cite{mie_beitrage_1908} for a monochromatic incident wave. Hence, it can be expressed directly in the frequency domain.\\
Formally writing the RHS of Eq.~\ref{eq_scatter_wave_disp_TF} as FT and using the convolution theorem, one can see the transient of the scattering in a dispersive medium as the time convolution product of three terms:
\begin{equation}
\bm{\psi}^s(t,\theta,L) =g(t,-L) * d(t,L) * \bm{\phi}(t,\theta)
\label{eq_scatter_wave_disp_convol}
\end{equation}
where $d(t,L)$ is the time signal of the dispersion operator.
The optical path is $-L$ for the pulse term because the origin of the coordinates system is the scatterer center.
The two first terms of the RHS can be merged in $g'$ to represent the time signal of the incident chirped pulse:
\begin{equation}
\bm{\psi}^s(t,\theta,L) =g'(t,L) * \bm{\phi}(t,\theta)
\label{eq_singlescatter_wave_disp_convol_gprime}
\end{equation}
The last factor of the RHS of Eqs.~\ref{eq_scatter_wave_disp_convol} and \ref{eq_singlescatter_wave_disp_convol_gprime} is the impulse response function of the scatterer. 
It corresponds to the scatterer response to an infinitely short light pulse, \ie a Dirac delta function.
Since it plays a particular role in the following we drop the letter $\psi$ and label it $\phi$ instead. It is expressed as:
\begin{equation}
\bm{\phi}(t,\theta) =  \mathscr{F}^{-1}[ \bm{\psi}^{cw}(\omega,\theta)]
=\frac{1}{\sqrt{2 \pi}} \int_{-\infty}^{\infty}  \bm{\psi}^{cw}(\omega,\theta) \, \text{e}^{i \omega t} \, \mathrm d \omega
\label{eq_SIRF}
\end{equation}
In the following $\bm{\phi}(t,\theta)$ is referred to as the Scattering Impulse Response Function (SIRF).
Note that the original solution of the LMT derived by Mie \cite{mie_beitrage_1908} can be seen from the same viewpoint. Indeed, the temporal response $\bm{s}_{\omega_0}(t)$ of a scatterer illuminated by a monochromatic light source of frequency $\omega_0$ is the convolution of the SIRF by a pure sine function, which in the frequency domain is equal to the Dirac delta  function. Hence:
\begin{equation}
\bm{s}_{\omega_0}(t,\theta) =  \frac{1}{\sqrt{2 \pi}} \int_{-\infty}^{\infty}  \delta(\omega - \omega_0) \, \bm{\psi}^{cw}(\omega,\theta) \, \text{e}^{i \omega t} \, \mathrm d \omega
=  \bm{\psi}^{cw}(\omega_0,\theta) \, \text{e}^{i \omega_0 t}
\label{eq_SSIRF}
\end{equation}
Numerically speaking, it may not be possible to calculate the SIRF exactly. When $\omega \to \infty$, the parameter size $x$ also tends to infinity, which forbids any application of the LMT. The use of geometrical optics for large $x$ could be used, but not when the relative refractive index $m<1$ (\eg air bubble in water) where the geometrical optics approximation cannot be satisfied near the critical angle \cite{sentis_scattering_2016}.
To circumvent the exact computation of the SIRF, we smooth the delta function to allow numerical integration of Eq.~\ref{eq_SIRF}. It is referred to as the SSIRF (Smoothed SIRF) method in the following.
The SSIRF $\bm{\phi}_{S}$ is the response of the scatterer not to a Dirac delta function, but to a very short, finite pulse. Considering a virtual ultra short pulse as a Gaussian with a FWHM $\Delta t_v$ of a few electromagnetic wave cycles $2 \pi / \omega_0$:
\begin{equation}
\bm{\phi}_{S}(t,\theta) = \frac{\gamma}{2\sqrt{\pi}} \int_{\omega_{\min}}^{\omega_{\max}} \text{e}^{-[\gamma (\omega-\omega_0)/2]^2} \, \bm{\psi}^{cw}(\omega,\theta) \, \text{e}^{i \omega t} \, \mathrm d \omega
\label{eq_SSIRF_kernel}
\end{equation}
with $\gamma$ expressed by Eq.~\ref{eq_gamma} with $\Delta t_v$. In this case, the smoothed impulse response is a signal whose peak thickness is at least $\Delta t_v$, which decreases the temporal exactness of the SIRF. This is the inherent counterpart of smoothing the Dirac delta function.
There are limitations on the minimal $\Delta t_v$.
First, the pulse cannot be shorter than one cycle of the EM wave, otherwise the mean value of the pulse would be larger than zero, meaning that the Fourier spectrum admits a DC component.
In this case it would be necessary to propagate with the EM wave a transverse DC potential, constant in space, which is impossible.
This is known as the "zero-area" rule of a propagating pulse \cite{pollock_chapter_2008}.
Second, as for the SIRF, large frequencies lead to a large size parameter whose calculation is computationally expensive.
Third, in the case of dispersive media (both the propagation medium or the scatterer medium), it is necessary to know the refractive index over the whole variation range of the frequency.
\\Despite these limitations, the smoothed impulse response $\bm{\phi}_{S}(t,\theta)$ of the temporal scattering phase function allows to recover the physics of scattering, as validated later.
Also, as long as the zero-area rule is respected, the SSIRF is a time signal that corresponds to a realistic physical phenomenon, and therefore it can be used as standalone model for the scatterer response.\\
For illustration, the SSIRF is shown in Fig.~\ref{fig_illustre_SSIRF_time_signal} (left) as a map in the ($\theta,t$) space. The time signal corresponding to different angles is given on the right of the figure. The vertical lines superimposed on the map mark the direction where the time signals are plotted. Note that the time signal of the stripes at $t>$~\SI{0.8}{\pico\second} are not visible for angles $\ge 45^{\circ}$ because their relative intensity $I/I_{max}$ decreases below the limits of the figure (\dixmoins{12}).

\subsection{Transient of multiple scattering}

Decomposing the scatterer response into different terms (Eqs.~\ref{eq_scatter_wave_disp_TF}-\ref{eq_singlescatter_wave_disp_convol_gprime}) enables to decouple the sources that contribute to the final temporal signal. This is particularly useful when modeling multiple scattering.
The scattering of an ultrashort pulse by $N$ scatterers in a dispersive media can be formally expressed in the temporal domain by using Eq.~\ref{eq_scatter_wave_disp_convol}:
\begin{equation}
\bm{\psi}^s(t,\theta) =g(t,-L_0) *
\underbrace{d(t,L_0) * \bm{\phi}(t,\theta_0)}_{\text{1}^{\text{st}}\text{ event}}
*
\underbrace{d(t,L_1) * \bm{\phi}(t,\theta_1)}_{\text{2}^{\text{nd}}\text{ event}}
* ... *
\underbrace{d(t,L_{N-1}) * \bm{\phi}(t,\theta_{N-1})}_{\text{N}^{\text{th}}\text{ event}}
\label{eq_multiscatter_wave_disp}
\end{equation}
Using the convolution theorem one can express Eq.~\ref{eq_multiscatter_wave_disp} in the frequency domain:
\begin{equation}
\begin{aligned}
\mathcal{F} [\bm{\psi}^s](\omega,\theta) = G(\omega-\omega_0) & \times   \text{e}^{i (\omega T_0 - k L_0)} \, \bm{\psi}^{cw}(\omega,\theta_0) \\ 
& \times \text{e}^{i (\omega T_1 - k L_1)}  \, \bm{\psi}^{cw}(\omega,\theta_1) \\
... \\
& \times \text{e}^{i (\omega T_{N-1} - k L_{N-1})} \, \bm{\psi}^{cw}(\omega,\theta_{N-1})
\label{eq_multiscatter_wave_disp_TF}
\end{aligned}
\end{equation}
where $T_i$ and $L_i$ are respectively the propagation time and the optical path length between events $i-1$ and $i$, for $i>0$; and $\theta_i$ is the scattering angle for the event $i$. The angle $\theta$ in the LHS is the summation of all scattering angles $\theta = \sum \theta_i$.
By gathering the terms, we obtain:
\begin{equation}
\mathcal{F} [\bm{\psi}^s](\omega,\theta,L) =
G(\omega-\omega_0)
\text{e}^{i (\omega T_t - k L_t)}
\prod_i \bm{\psi}^{cw}(\omega,\theta_i) 
\label{eq_multi_scatter_wave_disp_TF}
\end{equation}
where ($T_t,L_t$) = ($\sum{T_i},\sum{L_i}$). Equation~\ref{eq_multi_scatter_wave_disp_TF}, as Eq.~\ref{eq_scatter_wave_disp_TF}, can be regarded as a product of three terms corresponding to (i) the non-chripded pulse from the laser, (ii) the dispersion term corresponding to the whole optical path from the laser to the last scatterer and (iii) the scattering term repeated $N$ times for different angles.
When inverted into the time domain, Eq.~\ref{eq_multi_scatter_wave_disp_TF} is the time signal on a virtual detector located on the last scatterer surface and recording the far field EM wave in the polar $\theta$ direction at time $t$. It yields:

\begin{equation}
\bm{\psi}^s(t,\theta, L_t) =g(t,L_t) * d(t,L_t) *
\bigotimes_i \bm{\phi}(t,\theta_i)
\label{eq_multiscatter_wave_disp_convol}
\end{equation}
where the large dyadic symbol ${\otimes}$ represents multiple convolutions.
Equation \ref{eq_multiscatter_wave_disp_convol} can be written down:

\begin{equation}
\bm{\psi}^s(t,\theta, L_t) =g'(t,L_t) *
\bigotimes_i \bm{\phi}(t,\theta_i)
\label{eq_multiscatter_wave_disp_convol_gprime}
\end{equation}
where the first term represents the pulse spreading due to dispersion and the second term represents scattering.
Note that Eqs.~\ref{eq_multiscatter_wave_disp_convol} and \ref{eq_multiscatter_wave_disp_convol_gprime} refer to only one photon and one trajectory. Indeed, the total optical path length $L_t$ inside the dispersive medium depends on the series of scattering angles ($\theta_0...\theta_{N-1}$) which define a possible way to reach the $N^{th}$ scatterer from the laser source. The same comments goes to the time signal $\bigotimes_i \bm{\phi}(t,\theta_i)$.
Hence Eqs.~\ref{eq_multiscatter_wave_disp_convol} and \ref{eq_multiscatter_wave_disp_convol_gprime} must be integrated over all the possible scenarii represented by ($\theta_0...\theta_{N-1}$). This is discussed in the next section. Also, note that $L_t$ is the total path length of the photon inside the dispersive medium and does not include the path length inside the scatterer.
\\
In the case of particles of different diameters Eqs.~\ref{eq_multi_scatter_wave_disp_TF} and \ref{eq_multiscatter_wave_disp_convol_gprime} yield:

\begin{subequations}
\begin{align}
\mathcal{F} [\bm{\psi}^s](\omega,\theta,L_t) = & \ 
G(\omega-\omega_0)
\text{e}^{i (\omega T_t - k L_t)}
\prod_i \bm{\psi}_{d_i}^{cw}(\omega,\theta_i)\\
\bm{\psi}^s(t,\theta,L_t) =  & \  g'(t,L_t) *
\bigotimes_i \bm{\phi}_{d_i}(t,\theta_i)
\label{eqs_multiscatter_wave_disp_convol_gprime_polydisp}
\end{align}
\end{subequations}
where the subscript $d_i$ indicates a scattering event by a particle of diameter $d_i$.

\section{Temporal Monte Carlo techniques\label{sec_montecarlo}}

\subsection{Solving the Radiative Transfer Equation}

The Monte Carlo (MC) technique is a stochastic method to solve the deterministic RTE given by:
\begin{equation}
\frac{1}{c} \frac{\partial}{\partial t} I(\bm{z},\bm{\Omega},t) + \frac{\partial}{\partial z} I(\bm{z},\bm{\Omega},t)
= -(k+s) I(\bm{z},\bm{\Omega},t) + s \int_{4 \pi} I(\bm{z},\bm{\Omega},t) f(\bm{\Omega}, \bm{\Omega}',t) \, \mathrm d \bm{\Omega}'
\label{eq_rte_equation_basics}
\end{equation}
where $I(\bm{z},\bm{\Omega},t)$ is the intensity at location $\bm{z}$ propagating in the direction $\bm{\Omega}$ at time $t$. The terms $k$ and $s$ are the absorption and scattering coefficients, respectively. The second term on the RHS represents the scattering in all directions where $\mathrm d \bm{\Omega}'$ is the elementary solid angle about the direction $\bm{\Omega}'$ and $f(\bm{\Omega}, \bm{\Omega}',t)$ is the time-dependent phase function.
The LHS is the total derivative in an Eulerian frame and can be written in the Lagrangian frame as $\mathrm d I(\bm{z},\bm{\Omega},t) / (c \, \mathrm dt)$.
Equation~\ref{eq_rte_equation_basics} is based on the assumption that interferences are negligible.
This is justified by two sets of assumptions. First, the wavelength is negligible versus the dimension of the domain and the mean inter scatterer distance. Second, in our configuration the occurrence of interference is further diminished because the scatterers have a large parameter size ($x>100$), thus promoting forward scattering, and (ii) the laser pulse is ultrashort, thus the wavefront is localized in space and time. These two reasons decrease the probability that different optical paths of the coherent wave front will cross each other and interfere. \\

When solving Eq.~\ref{eq_rte_equation_basics} with the MC method, each ray of light is represented by an energy quanta (referred to as “photon” for convenience) traveling in the turbid medium and carrying the same quantity of elementary energy. When a photon undergoes a scattering event, the scattering angle is randomly drawn according to a given Probability Density Function (PDF), which corresponds to the phase function.
This method can be regarded as converting the anisotropic redistribution of energy in different directions after the scattering to a probability to have a photon with the same directions.
As a particle method, the MC method is very versatile and the whole history of the particle can be accounted for such as the trajectory, the number of scattering event, etc, which allows a deep insight in the scattering phenomenon.
This approach was used to model transient of multiple scattering based on the photon time of flight only \cite{berrocal_laser_2007, linne_ballistic_2009}, or with accounting the transient inside large scatterers \cite{roze_interaction_2003, calba_monte_2006, calba_ultrashort_2008, calba_interaction_2008}.

\subsection{Energy transport approximation}

To make the link between the radiative transport theory that describes power and intensity, and the transient of scattering that describes amplitudes of electromagnetic waves, it is necessary to express the power of the scattering transient. Therefore, we consider the signal of the intensity $I(t,\theta)$ recorded by a time detector capable of resolving different directions. It is found by taking the squared modulus of Eq.~\ref{eqs_multiscatter_wave_disp_convol_gprime_polydisp}: 
\begin{equation}
I(t,\theta,L_t) = | \bm{\psi}^s(t,\theta) |^2 =  \left| g'(t,L_t) *
\bigotimes_i \bm{\phi}_{d_i}(t,\theta_i) \right|^2
\label{eq_power_of_transient_exact}
\end{equation}
Although exact, Eq.~\ref{eq_power_of_transient_exact} is not practical to transport particles of energy in the turbid media because the convolutions are applied to the EM wave amplitudes.
Therefore, we assume that we can distribute the modulus ($|\cdot|^2$) inside the convolutions, \ie that the energy of the photon follows the same time distribution as its complex amplitude:

\begin{equation}
I(t,\theta,L_t)  \approx  \left| g'(t,L_t)  \right|^2 *
\bigotimes_i \left|  \bm{\phi}_{d_i}(t,\theta_i) \right|^2
\label{eq_energy_approx}
\end{equation}
Again, this approximation is justified by the fact that interference can be neglected. Equation~\ref{eq_energy_approx} is rewritten in terms of intensity:

\begin{equation}
I(t,\theta,L_t)  \approx  I_p(t,L_t)  *
I_{\phi,d_0}(t,\theta_0) *
I_{\phi,d_1}(t,\theta_1) *
... *
I_{\phi,d_{N-1}}(t,\theta_{N-1})
\label{eq_multiscat_energy_approx}
\end{equation}
where $ I_p(t,L_t) = \left| g'(t,L_t)  \right|^2 $ is the intensity of the incident pulse after a propagation on distance $L_t$.
The sequential convolution in this equation allows us to model the scattered intensity collected by a detector with a PDF, as shown in the next section.

\subsection{Two modeling approaches\label{ssec_two_approaches}}

\begin{figure}[!htb]
	\begin{center}
	\def \svgwidth {0.8\textwidth}
	{\scriptsize
	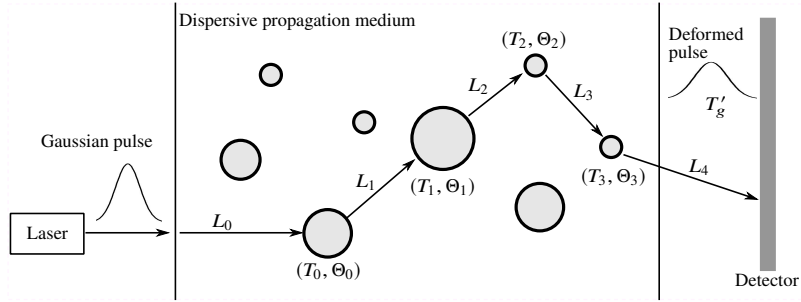
	}
	\caption{2D Sketch of the pure Monte Carlo approach.}
	\label{fig_sketch_m1}
	\end{center}
\end{figure}

\begin{figure}[!htb]
	\centering
	\includegraphics[width=\columnwidth,keepaspectratio]{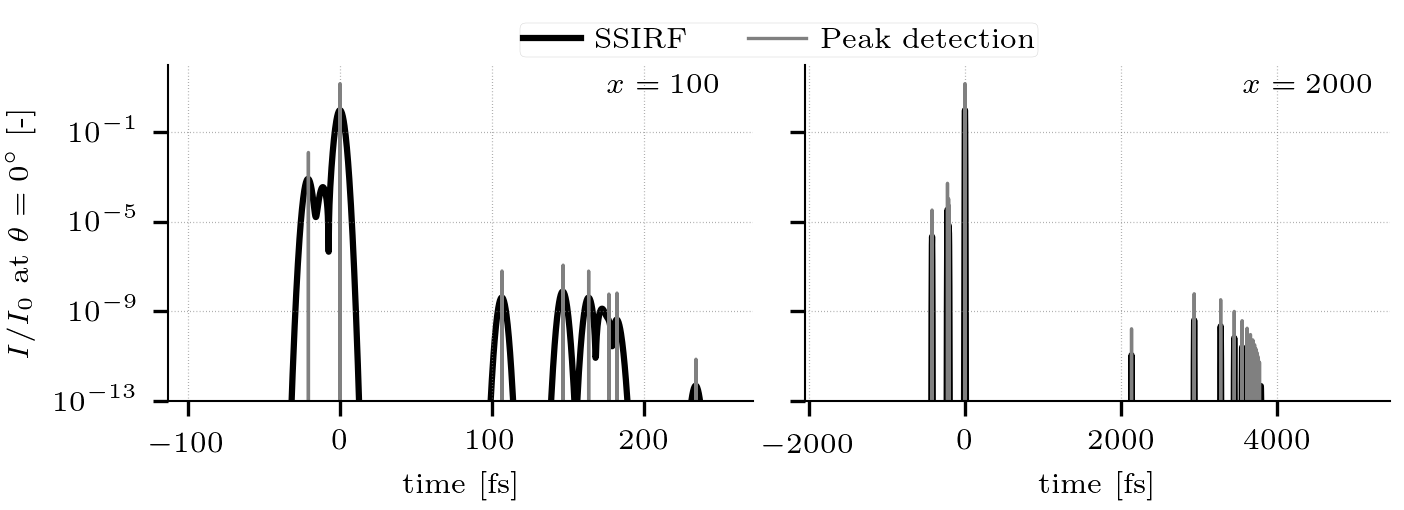}
	\caption{Time signal at $\theta=0^{\circ}$ for a bubble in water of size parameter 100 (left) and 2000 (right). Black curve is the SSIRF, grey curve are the detected peaks.}
	\label{fig_illustre_two_methods_time_sig_peaks}
\end{figure}

\begin{figure}[!htb]
	\begin{center}
	\def \svgwidth {0.8\textwidth}
	{\scriptsize
	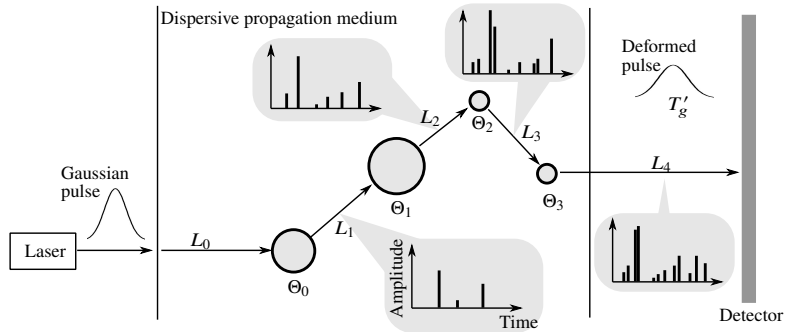
	}
	\caption{Sketch of the Comb method.}
	\label{fig_sketch_m2}
	\end{center}
\end{figure}

In the rest of this study we will compare two different methods to compute the scattering delay in single or multiple scatterings. Both methods are formulated with the energy approximation although they would be also valid when transporting the amplitude (and therefore considering the polarization). The first method is a pure Monte Carlo method, and referred to as Method~1.
For each scattering event, the delay is randomly drawn according to a given multivariate PDF depending on time and direction. In case of multiple scattering, the total delay due to scattering is the sum of the delay of each event.
Let us consider $I(t,\theta)$ in Eq.~\ref{eq_multiscat_energy_approx} as the time signal on a detector able to separate the different incident angles.
$I(t,\theta)$ can be regarded as a multivariate PDF to determine $(t,\theta)$.
Since $I(t,\theta)$ is expressed as multiple products of convolution, we can benefit from the fact that the random variable described by a convolution of two PDFs is equal to the sum of the random variables described by the two PDFs.
Let us define the random variable $T_{g'}$ as the "in-time" position of the photon in the pulse, and associate it to the univariate PDF given by $I_p(t,L)$. In the same manner we define the random variables $(T_{i},\Theta_{i})$ as the time and direction of the scattered photon after the $i$\textsuperscript{th} event and we associate them to the multivariate PDF $I_{\phi,d_i}(t,\theta)$. Therefore, the random variables $T_\text{detector}$ and $\Theta_\text{detector}$, respectively defined as the time and direction of the photon impacting the detector are written:

\begin{subequations}
\begin{align}
\Theta_\text{detector} = & \ \Theta_0 + \Theta_1 +  ... + \Theta_{N-1} \\
T_\text{detector} = & \ T_{g'} |_L +  T_0 |_{\Theta_0}+ T_1|_{\Theta_1} + ... + T_{N-1}|_{\Theta_{N-1}} \label{eq_fullMC_time} 
\end{align}
\label{eq_random_var_meth1_multi}
\end{subequations}
where $N$ is the number of scattering events. Equation~\ref{eq_fullMC_time} shows conditional probability with $T_{g'} |_L$ being the random variable of the pulse delay given a optical path length $L$ and $T_i |_{\Theta_i}$ being the univariate random variable on scattering time in the polar direction $\Theta_i$.
We use the SSIRF as the multivariate PDF, hence in Eq.~\ref{eq_multiscat_energy_approx} $I_{\phi,d_i}(t,\theta) = | \bm{\phi}_S(t,\theta) |^2$ with $\bm{\phi}_S(t,\theta)$ expressed by Eq.~\ref{eq_SSIRF_kernel} for a given diameter $d_i$. Note that this is physically consistent because the SSIRF is computed over at least one cycle of the electromagnetic wave and hence it respects the zero area rule. This means that the PDF depicts a physical phenomenon and a random generator based on this PDF renders a physically consistent stochastic signal.
\\
The algorithm of Method~1 is sequentially described as follows and illustrated in Fig.~\ref{fig_sketch_m1}. 
A numerical photon is shot at the left inlet boundary of the slab. At each scattering event, first the polar angle of scattering is randomly drawn according to the marginal univariate Cumulative Density Function (CDF) $F(\theta)$ defined between 0 and $\pi$ by:
\begin{equation}
F(\theta) = \int_{0}^{\theta} f(u) \, \sin(u) \, \mathrm d u
\quad \text{with} \quad
f(u) = \int_{t_{\min}}^{t_{\max}} I_{\phi}(t,u) \, \mathrm d t
\end{equation}
In other words $F(\theta)$ represents the polar distribution of the energy integrated in time when a short pulse is scattered by a spherical particle. It is different from the steady state scattering phase function because it filters out potential interferences between different refraction modes \cite{mees_time-resolved_2001, chaussonnet_scattering_2020}.
Since we consider only the energy, we neglect the polarization information and hence the azimuthal scattering angle $\Phi_i$ is equally distributed between 0 and $2 \pi$.
Once $\Theta_i$ is selected, we randomly draw the scattering delay $T_i|_{\Theta_i}$ according to the CDF $H(t,\Theta_i) = \int_{t_{\min}}^{t} I(t',\Theta_i) \, \mathrm d t'$, which is added to the total scattering time later detected.
After this ($i$\textsuperscript{th}) event the photon is deflected in the direction given by $(\Theta_i,\Phi_i)$.
As detailed below, the distance between two scattering events, and the scatterer diameter are randomly drawn according to an exponential distribution, and a prescribed droplet size distribution, respectively.
When the photon exits the slab and reaches the detector, the total optical path is converted into the time necessary to travel this distance. This is the time of flight of the photon. In case of the photon flying through a dispersive medium, the pulse spread is taken into account by considering the optical path inside the dispersive medium. The ``in space'' spread is converted into ``in time'' spread, which can be considered at first order as increasing the variance of the initial Gaussian pulse, provided that the pulse has still a Gaussian profile. This pulse is then expressed in terms of CDF by the error function of zero mean and $\sigma$ standard deviation as $I_p(T_{g'},L) \propto \text{erf}(T_{g'},0,\sigma(L))$. Also note that $L$ in  $I_p(T_{g'},L)$ is an input parameter of the CDF.
From an algorithmic point of view, this method is easy to implement but requires to load the map $I(t,\theta)$ for each diameter of the particles. Also, it requires a very large amount of photons to converge, as seen later.
\\\\
The second method was originally proposed by Calba \etal \cite{calba_monte_2006}. After a scattering event, the numerical photon does not only transport one single scattering delay, but the whole time signal such as $I_{\phi,d_0}(t,\Theta_0)$  in Eq.~\ref{eq_multiscat_energy_approx} for the first scattering event. After a second scattering event, the time signal is the convolution of the two time signals $I_{\phi,d_0}(t,\Theta_0) * I_{\phi,d_1}(t,\Theta_1)$, and so on. 
When the photon impact the detector, its time signal is convoluted with the chirped pulse signal ($I_p(t,L)$). Time signals of all photons reaching the detector are summed together.
To decrease the memory requirement of such a procedure, we idealize the scattering time signal as a train of peaks, \ie a comb, as done in Calba \etal \cite{calba_monte_2006}. This idealization is based on the fact that when the pulse width ($=c \Delta t$)
is much smaller than the particle diameter, it acts as a scan in time and angle of the particle (see \cite{chaussonnet_scattering_2020}), thus leading to a time signal made of very thin peaks. By virtually reducing the pulse duration to a Dirac delta function, it is assumed that the time signal reduces to a Dirac comb where peaks are not regularly distributed in time. This method could thus be coined the Comb Transport, but is referred to as Method~2 in the following.
We are aware that this assumption is more a naive geometrical idealization than a real physical argument. One could argue, as mentioned above, that the zero area rule must be fulfilled and hence the peaks in the time signal could not be thinner than $2 \pi / \omega_0$. However, because the resulting comb is eventually convolved by the chirped Gaussian pulse
(\eg $I_p(t,z_{\text{detector}})$ in Eq.~\ref{eq_multiscat_energy_approx}), it is trimmed in the frequency domain to the frequency range of the pulse,
thus unphysical large frequencies are filtered out. 
Hence its final time representation is equal to the one of the pulse $> 2 \pi / \omega_0$.\\
The comb signal is generated by selecting the peaks of the SSIRF time signal. It is to be noted that when the time signal is noisy (for instance when the SSIRF is computed with a varying refractive index as seen later), it can be necessary to smooth the time signal. The results of the peak detection is illustrated in Fig.~\ref{fig_illustre_two_methods_time_sig_peaks}.
The black line represents the intensity of the time signal at forward scattering ($\theta=0^{\circ}$) for a droplet in air of size parameter of 100 (left) and 2000 (right) illuminated by a laser ($\lambda_0=$~\SI{600}{\nano\meter}) virtual pulse of two EM cycles. The grey segments are the detected peaks.
For small particle ($x=100$), the virtual pulse width becomes closer to the particle diameter and the peak identification process misses some peaks. On the contrary for large scatterers the time signal is a set of individual and well separated peaks, easing the detection process.
\\The sequence of Method~2 is illustrated in Fig.~\ref{fig_sketch_m2}.
During a scattering event, the polar angle of scattering is randomly drawn as in Method~1. The angle of scattering, to which corresponds a given comb, is recorded in a list. When the photon finally impacts the detector, the list of scattering angles is converted into a list of combs, which are convoluted recursively.
A threshold on the amplitude is applied to dismiss weak peaks.
The resulting comb is eventually convolved with the time signal of the chirped pulse.
\\Method~2 has the advantages of being much lighter in memory because one peak is determined by its amplitude and its time and hence requires only two scalars per peak. Also, the discrete convolution of a comb is easy to implement with recursion and easy to optimize by sorting the peaks by decreasing amplitude. The results of the convolution is that amplitudes are multiplied and the times are added. The interested reader is referred to \cite{calba_monte_2006} for a detailed explanation on the Comb Transport.\\
The results of the two methods will be compared to each other and to other results from literature in Section~\ref{sec_transient_multi_energy}.

\subsection{Monte Carlo code: Scatter3D}

The Monte Carlo simulations were performed with the code Scatter3D \cite{jenny_computing_2007}, which was developed to solve the RTE (Eq.~\ref{eq_rte_equation_basics}) in steady state. Scatter3D can handle advanced optic features such as fluorescence \cite{sormaz_stochastic_2009}, polarization \cite{sormaz_stochastic_2010}, birefringence \cite{sormaz_influence_2010} or time correlation \cite{sormaz_breakdown_2014}. Scatter3D was applied to characterize bioorganic tissues \cite{sormaz_influence_2010} or to predict halftone patterns of ink coverage \cite{sormaz_predicting_2008}. One of the particularities of Scatter3D is that it uses a stencil approach where scattering events are not resolved one by one, but as a group over a given stencil. This is achieved by precomputing all scattering events for a collection of different scenarii in an elementary volume, and later apply these possible scenarii. Although this approach can speed up simulations tremendously, this feature was not used here.\\
The distance between two scattering events $l_{\text{scat}}$ is drawn from the memoryless exponential law $e^{-l_{\text{scat}}/l_m}$, where $l_m$ is the mean free path length of the photon. The optical thickness $\tau$ of the medium is given by $\tau = l_s / l_m$ where $l_s$ is the geometrical thickness of the slab.\\
The two methods to compute the scattering transient presented in Section~\ref{ssec_two_approaches} were incorporated in the code.
The SSIRF are precomputed by a Python script and processed to provide the $(t,\theta)$ map for Method~1 and the comb signal for Method~2. These data are then loaded at the initialization of Scatter3D.
As polydisperse clouds of droplets (or bubbles) are simulated in the following, it is necessary to precompute the SSIRF for different diameters and load them in Scatter3D.
During the simulation, when a photon impact a scatterer, a diameter is randomly draw from the pool of precomputed diameters, according to their size distribution.

\section{Spreading of the pulse in water \label{sec_pulse_deformation}}

\begin{figure}[!htb]
	\centering
	\includegraphics[width=0.495\columnwidth,keepaspectratio]{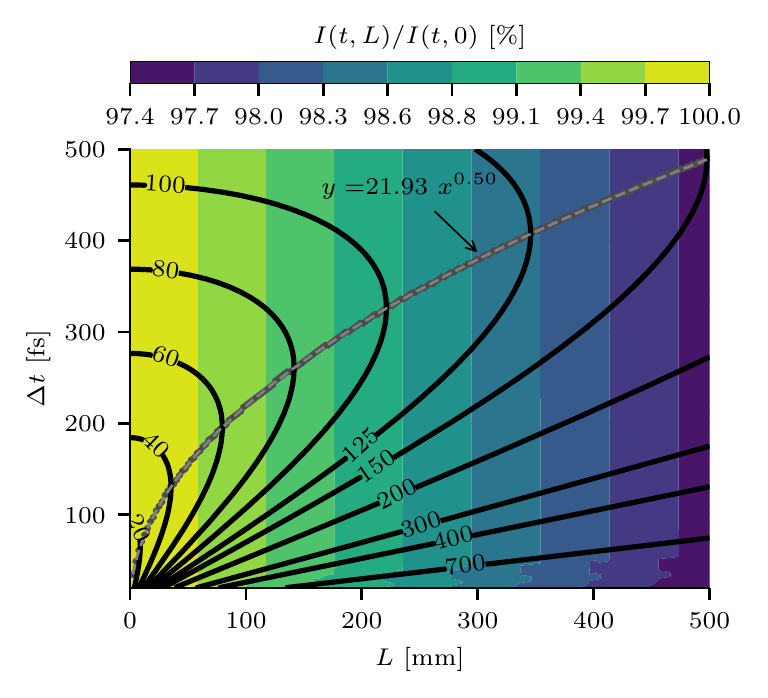}
	\includegraphics[width=0.495\columnwidth,keepaspectratio]{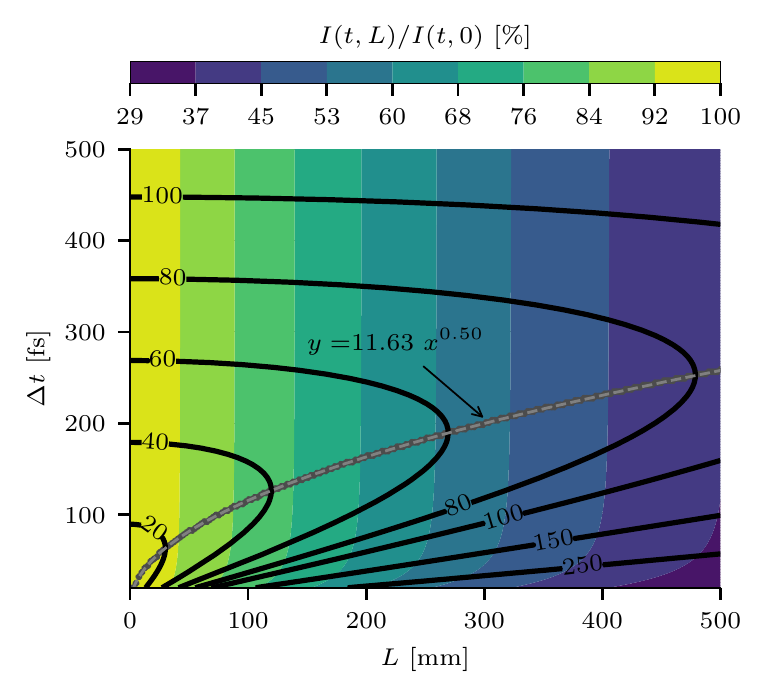}
	\caption{Map of intensity and isocontours of $\Delta s$ in micrometer versus the pulse duration $\Delta t$ and the distance to laser $L$ for $\lambda_0=400$~nm (left) and 800~nm (right).}
	\label{fig_isolines_ds_vs_L0_dt}
\end{figure}

Water is a dispersive optical media and it is important to optimize for pulse length and carrier frequency to minimize pulse chirping.  Absorption by water is also taken into account.
Because of the pulse spectrum broadening, the spreading depends also on the pulse duration. This is illustrated in the Appendix~\ref{appendix_pulsespread}.
To quantify the spreading in various conditions, the propagation of a short pulse in water is simulated according to Eq.~\ref{eq_incident_wave_disp_new_frame} for different pulse duration $\Delta t$ and different distance to source $L$. They are summarized in Table~\ref{tab_ref_case}.
We quantify the pulse spread in space by the FWHM of its extension, $\Delta s$.
\begin{table}[!htb]
	\centering
	\caption{Operating parameters for the spreading of the pulse.}
	\label{tab_ref_case}
	\begin{tabular}{  l l c }
	\hline
	$\Delta t$ & [fs] & 20 - 500 \\
	$L$ & [mm] & 0 - 500 \\
	\hline
	\end{tabular}
\end{table}
\\In Fig.\ref{fig_isolines_ds_vs_L0_dt} the isolines show the contour of $\Delta s$, while the colormaps show the extinction, for different pulse duration $\Delta t$ at different depth $L$ in water, for a carrier wavelength of 400 (left) and \SI{800}{\nano\meter} (right).
The grey dots mark the minimal pulse width for at a given $L$ found numerically, and the grey line is the corresponding fitting curve. Since they only shows dispersion, these isolines are related to the real part of the refractive index.
The pulse is much more spread at $\lambda_0=$~\SI{400}{\nano\meter} than at \SI{800}{\nano\meter}. This is because the slope magnitude of the real part of the refractive index $(\partial n / \partial \lambda)|_{\lambda_0}$ at $\lambda_0$~=~\SI{400}{\nano\meter} is much larger (\SI{-95}{\per\milli\meter}) than that at $\lambda_0$~=~\SI{800}{\nano\meter} (\SI{-17}{\per\milli\meter}), which leads to a larger spreading of the pulse.
The key element of Fig.~\ref{fig_isolines_ds_vs_L0_dt} is that there is an optimal pulse duration to minimize the spread of the pulse at a given distance.
Another interesting point from Fig.~\ref{fig_isolines_ds_vs_L0_dt} is the exponent 0.50 of the fitting correlations, suggesting that the optimal $\Delta t$ to minimize $\Delta s$ is proportional to $\sqrt{L}$. This highlights the fact that in the current conditions, the real part of the refractive index is very well represented by a second order Taylor expansion in $1/\lambda$.
This is demonstrated in the Appendix~A.%
\\\\
The maps of extinction here represent the total energy of the pulse normalized by its value at $z=0$ such as $I(L)/I(0)$ where $I(L) = \int | \bm{X}(t,z=L)|^2 \, \mathrm d t$.
The much weaker extinction at \SI{400}{\nano\meter} compared to that at \SI{800}{\nano\meter} is strongly illustrated with a normalized intensity larger than 97\% at 500 mm whereas the pulse intensity is reduced by two third at 500 mm for $\lambda_0$~=~\SI{800}{\nano\meter}.
For very short pulses ($<$\SI{70}{\femto\second}) at $\lambda_0$~=~\SI{400}{\nano\meter}, the extinction shows non-monotonic variations versus $\Delta t$. This is because the imaginary part of refractive index reaches a minimum at $\lambda \approx$~\SI{475}{\nano\meter} as shown in Fig.~\ref{fig_n_water_versus_lambda_alloverlaid}. For very short pulses (\ie \SI{20}{\femto\second}), the color broadband leads to frequency up to \SI{520}{\nano\meter} on the other side of the minimum. But for longer pulses, the maximum wavelength reaches the minimum of the extinction, and hence the overall extinction is smaller.
\\\\
To conclude, when a short ($<$~\SI{500}{\femto\second}) pulse of visible light propagates in water, the broadband color leads to its spread due to dispersion, and to its weakening due to extinction. It is not possible to minimize the pulse spreading and its extinction at the same time. At large wavelengths (\SI{800}{\nano\meter}), the dispersion effects are lower but the extinction is large, whereas small wavelengths (\SI{400}{\nano\meter}) are strongly dispersed but very weakly attenuated. However if the water depth is known, maps in Fig.~\ref{fig_isolines_ds_vs_L0_dt} are useful to determine the optimal pulse duration. Another solution could be to chirp the pulse before it enters the dispersive medium to partly pre-compensate the dispersion.

\section{Sensitivity of the temporal phase function to numerical parameters\label{sec_SSIRF}}

\begin{figure}[!htb]
	\centering
	\includegraphics[width=0.9\columnwidth,keepaspectratio]{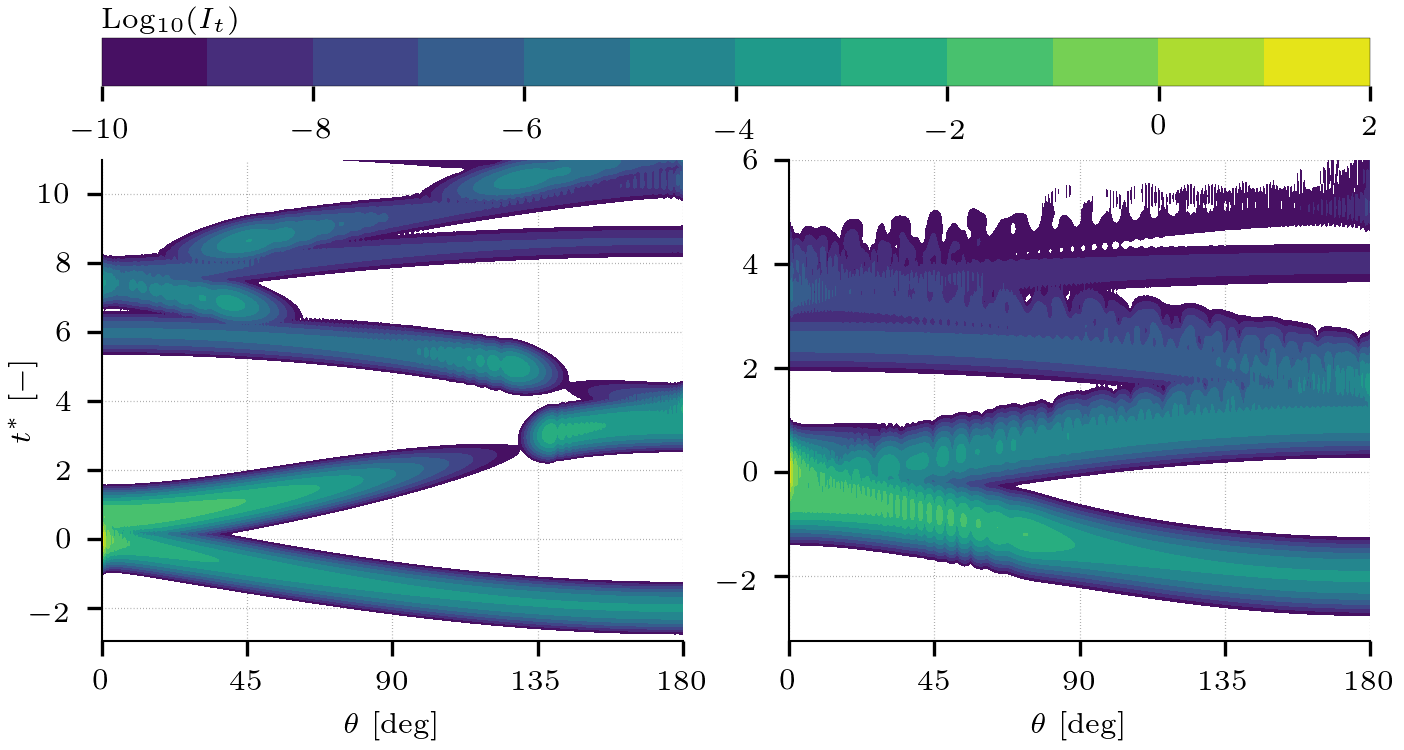}
	\caption{Transient scattering phase function ($x=500$) for a water droplet in air (left) and air bubble in water (right) illuminated by a 100 fs and 800 nm laser pulse.}
	\label{fig_transient_scattering_illustration}
\end{figure}

In this section, we investigate the numerical parameters to accurately compute the scattering transient of the real pulse and of the SSIRF, the objective being to find the optimal parameters to retrieve Eq.~\ref{eq_singlescatter_wave_disp_convol_gprime}.
As a first step we assume a non-chirped pulse (non-dispersive medium), so that Eq.~\ref{eq_singlescatter_wave_disp_convol_gprime} simplifies to:
\begin{equation}
\bm{\psi}^s(t,\theta) =g(t) * \bm{\phi}(t,\theta)
\label{eq_singlescatter_wave_disp_convol_gprime_recall}
\end{equation}
First we will examine the sensitivity of the LHS to the frequency resolution, then we will use the LHS as a reference and examine the sensitivity of $\bm{\phi}(t,\theta)$ to satisfy Eq.~\ref{eq_singlescatter_wave_disp_convol_gprime_recall}.\\\\
The LHS of Eq.~\ref{eq_singlescatter_wave_disp_convol_gprime_recall} is computed with Eq.~\ref{eq_scatter_wave_disp} for $L=0$ and stands as the reference.
The reference case consists of a particle of $x=500$ size parameter. The FWHM in time of the pulse is $\Delta t$~=~\SI{100}{\femto\second} and the wavelength in vacuum of the carrier is $\lambda_0$~=~\SI{800}{\nano\meter}. Two types of configurations are studied: (i) a water droplet in air and (ii) an air bubble in water, corresponding to a relative refractive index larger and lower than one.
The corresponding temporal scattering phase functions are illustrated in Fig.~\ref{fig_transient_scattering_illustration}, where the $y$-coordinate is expressed in terms of reduced time $t^*=t c_0 / (n_{pm} r_s)$.
The different modes of reflection, refraction, and internal reflections are discernible even though their overlap. For detailed comments on these maps, the reader should refer to \cite{chaussonnet_scattering_2020}.

\subsection{Time discretization  and frequency resolution}

\begin{figure}[!htb]
	\centering
	\includegraphics[width=\columnwidth,keepaspectratio]{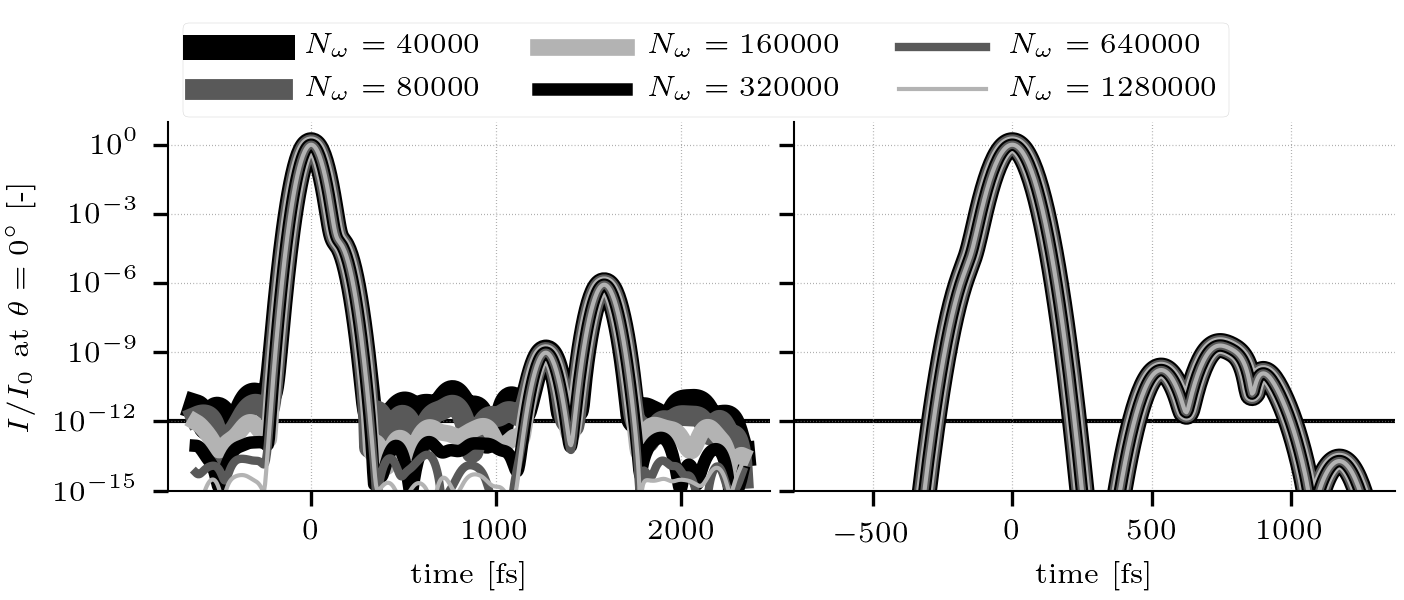}
	\caption{Time signal at $\theta=0^{\circ}$ for a water droplet in air (left) and air bubble in water (right) illuminated by a 100 fs and 800 nm laser pulse ($x=500$) computed with various frequency resolutions.}
	\label{fig_transient_scattering_dt100_x500_influenceNnu}
\end{figure}

In the frequency domain, we discretize the frequency axis between $\omega_{\text{min}}$ and $\omega_{\text{max}}$ where the relative peak of Fourier Transform of the pulse ($G/G(0)$ in Eq.~\ref{eq_pulse_FT_pulse}) is larger than a precision criterion $\epsilon$:
\begin{equation}
\omega_{\text{min}}, \omega_{\text{max}} = \omega_0 \pm \Delta \omega / 2
\quad \text{where} \quad
\Delta \omega = \frac{\sqrt{2\log (1/ \epsilon)}}{\gamma}
\label{eq_Delta_omega}
\end{equation}
where $\gamma$ is the time constant such that the temporal signal of the pulse is $g(t,z=0) = \e^{-(t/\gamma)^2}$ as given by Eq.~\ref{eq_gamma}.
Please note the difference of definition between $\Delta t$ and $\Delta \omega$, the former being the FWHM \ie Eq.~\ref{eq_Delta_omega} with a precision criterion of $\epsilon=0.5$ whereas the latter is the full width for a precision criterion $\epsilon \to 0$.
The temporal resolution $\mathrm dt$ of the signal given by the inverse Fourier Transform is:
\begin{equation}
\mathrm dt = \frac{2 \pi}{\omega_{\max} - \omega_{\min}} = \frac{2 \pi}{\Delta \omega} = \frac{\pi \Delta t } {4 \sqrt{\log(2) \log(1/ \epsilon)}}
\end{equation}
With $\epsilon=$~\SI{e-30}, $\mathrm dt \approx \Delta t / 8.81$, which means that the largest half of the pulse is resolved by 9 samples. This is considered here as too coarse, and thus $\Delta \omega = \omega_{\max} - \omega_{\min}$ will be extended by zero padding (and keeping $\mathrm d \omega$ constant) in the frequency domain to impose $\mathrm dt \approx \Delta t /20$.\\
The resolution $\mathrm d \omega$ of the frequency axis is set by the number of sample $N_{\omega}$ between $\omega_{\min}$ and $\omega_{\max}$,
so that $\mathrm d \omega = \sqrt{2\log (1/ \epsilon)} / (N_{\omega}\gamma)$.
Therefore, $\mathrm d \omega$, and hence the accuracy of the transient scattering phase function, depend on both the frequency resolution and on the pulse duration, due to the presence of $\gamma$.
Its influence is discussed in the following. After preliminary tests, it was found that the scattering angle where spurious modes are the more prominent is 0°. Thus we limit our parameter study to this angle.
\\\\
First, we check the influence of the frequency resolution on the real (\SI{100}{\femto\second}, \SI{800}{\nano\meter}) pulse.
Figure~\ref{fig_transient_scattering_dt100_x500_influenceNnu} shows the time signal of the scattered intensity $|\bm{\psi}^s(t,\theta=0^{\circ})|^2$ for a droplet (left) and a bubble (right) illuminated by a real pulse of non-polarized light with different frequency resolutions. For the droplet, the background noise is made of spurious modes resulting from the discretization of the frequency axis, and decreases as the resolution increases. Concerning the physical modes (for $I/I_0 \gtrapprox 10^{-10}$), they are predicted with the same accuracy for all $N_{\omega}$.
The time signal of the bubble (Fig.~\ref{fig_transient_scattering_dt100_x500_influenceNnu} right) is independent of $N_{\omega}$, highlighting a fast convergence with the frequency resolution for the integration of Eq.~\ref{eq_scatter_wave_disp}. This suggests a smooth variation of the term $\bm{\psi}^{cw}(\omega,\theta)$ for a relative refractive index $m$ lower than one.
We consider arbitrarily the signal below $\approx 10^{-12}$ as background noise. The minimum $N_{\omega}$ to properly resolve the transient of this operating point is 160000 and 40000 for the droplet and the bubble, respectively. We will however see in the following that the background noise intensity depends also on other parameters.

\subsection{Computation of the SSIRF with a constant index}

First we investigate the SSIRF with a constant refractive index for water. The refractive index of air is always considered constant.
To compute the SSIRF, we resolve Eq.~\ref{eq_scatter_wave_disp} for $L=0$ for a virtual short pulse of FWHM $\Delta t_v$ of a few cycles of the carrier such as
$\Delta t_v$~=~$\tau_0/2$, $\tau_0$, and $2\tau_0$
where $\tau_0=\lambda_0/c_0$.
The frequency resolution is set to $N_{\omega}=$1280000 for all cases. Please note that in the case of the SSIRF, the convergence for $m<1$ is also faster than for $m>1$, as for the real scattering function (Fig.~\ref{fig_transient_scattering_dt100_x500_influenceNnu}). It is however not depicted here.
\begin{figure}[!htb]
	\centering
	\includegraphics[width=\columnwidth,keepaspectratio]{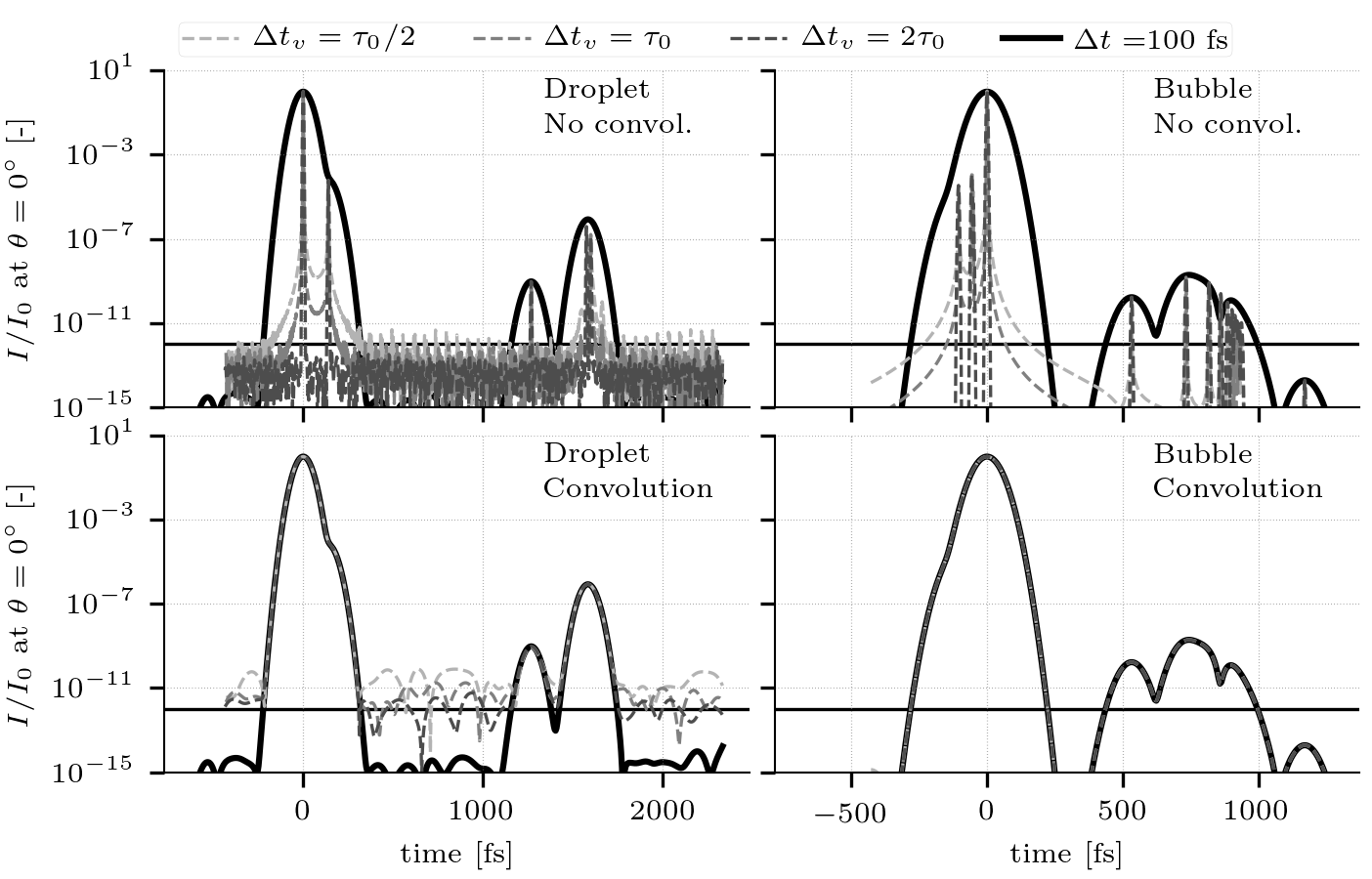}
	\caption{Time signal of the smoothed SIRF at $\theta=0^{\circ}$ for a water droplet in air (left) and air bubble in water (right) illuminated by a 800 nm laser pulse of duration $\tau_0/2$, $\tau_0$, $2\tau_0$, superimposed with the transient scattering function of the same particles illuminated by a $100$~fs pulse. Top: original SSIRF. Bottom: SSIRF convoluted the 100 fs pulse. Refractive indices are constant.}
	\label{fig_sSIRF_with_without_convol_x500_nodispersion}
\end{figure}
\\\\
All the results are summarized in Fig.~\ref{fig_sSIRF_with_without_convol_x500_nodispersion} whose layout is explained as follows. In the top row the intensities of the SSIRFs ($|\bm{\phi}(t,\theta=0^{\circ})|^2$) for different virtual pulse durations $\Delta t_v$ are shown for the droplet (left) and bubble (right). The transient scattered intensity for a real pulse ($\Delta t=$\SI{100}{\femto\second}) is given for comparison purpose in thick black line.
In the bottom row of Fig.~\ref{fig_sSIRF_with_without_convol_x500_nodispersion} the SSIRFs are convolved with an ideal Gaussian pulse of $\Delta t=$\SI{100}{\femto\second} and the resulting intensity is compared to the one of the real pulse.\\
First, we discuss the non-convolved SSIRF (Fig.~\ref{fig_sSIRF_with_without_convol_x500_nodispersion} top).
It is observed with the droplet configuration that the intensity of the spurious modes depends also on $\Delta t_v$ ($\equiv \gamma$) as mentioned above, because $\mathrm d \omega$ is inversely proportional to $\Delta t_v$.
Another striking effect is that for $\Delta t_v \le \tau_0$, the signal around the primary peaks (around $t$~=~\SI{0}{\femto\second}) raises above $10^{-12}$ in the approximate shape of a Laplace distribution with large tails. This is particularly visible for the bubble.
This effect is independent of $N_{\omega}$, and is attributed to a non-physical artifact such as the violation of the "zero area" rule when $\Delta t_v < \tau_0$.
Interestingly this effect has no influence on the convoluted signals (Fig.~\ref{fig_sSIRF_with_without_convol_x500_nodispersion} bottom), where the matching between the real signal and the convolved signal is perfect for the bubble, most presumably because of the frequency trimming.
For the convolved signal of the droplet, the agreement is excellent on the peaks, but the intensity of the spurious modes depends on $\Delta t_v$ and decreases when $\Delta t_v$ increases. The good agreement on Fig.~\ref{fig_sSIRF_with_without_convol_x500_nodispersion} (bottom) validates the present approach (Eqs.~\ref{eq_singlescatter_wave_disp_convol_gprime} and \ref{eq_singlescatter_wave_disp_convol_gprime_recall}).

\subsection{Computation of the SSIRF with dispersion}

\begin{figure}[!htb]
	\centering
	\includegraphics[width=\columnwidth,keepaspectratio]{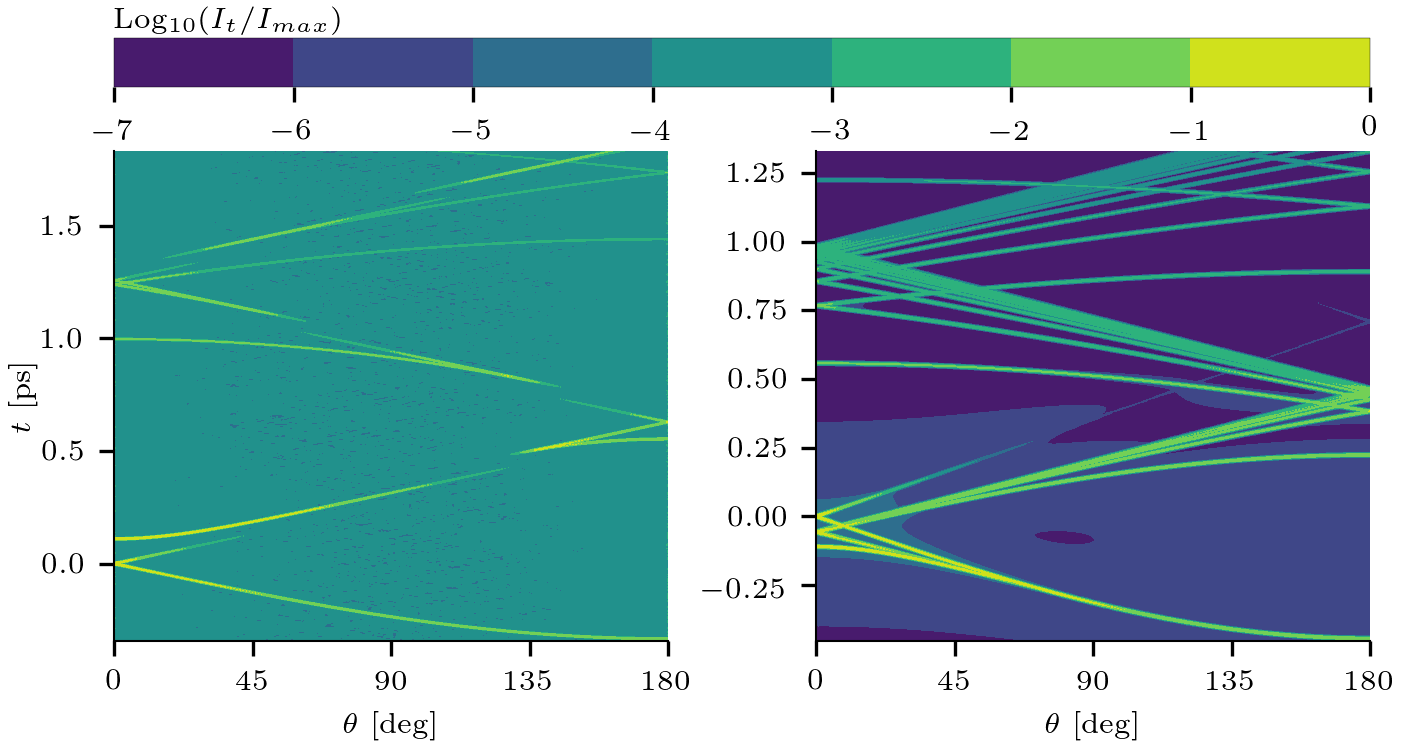}
	\caption{Map of the SSIRF for a scatterer of \SI{100}{\micro\meter} ($x \approx 500$). Droplet (left) and bubble (right).}
	\label{fig_illustration_SSIRF_map}
\end{figure}

In this part the dispersion and extinction are taken into account in the computation of the SSIRF. The scatterer is virtually located at the source ($L=0$), so that the pulse is not chirped, but the term $\bm{\psi}^{cw}(\omega,\theta)$, which depends on the relative refractive index has an additional dependency on $\omega$.
\\
In order to take the dispersion and extinction into account, it is necessary to know the variation of the refractive index over the whole range of wavelength. In the case of a real pulse, the broadband color is well included in the range where the index of water was accurately determined in the literature (see Fig.~\ref{fig_n_water_versus_lambda_alloverlaid}). However in the case of a virtual pulse, the broadband color is much larger with a wavelength ranging from \SI{148}{\nano\meter} to the infinity in the present conditions.
Hence, in this part we created a hybrid model by combining the databases from Harvey \etal \cite{harvey_revised_1998} and Segelstein \cite{segelstein_complex_1981} for the real part of the refractive index of water. This is motivated by the fact that the results from Segelstein \cite{segelstein_complex_1981} (i) deviate substantially from other more measurements of the literature \cite{austin_index_1976,schiebener_refractive_1990,mesenbrink_complex_1996} for visible light and (ii) are the only measurements for extreme wavelengths.
The details of this hybrid model are given in Appendix~B.%
\\All the results are gathered in Fig.~\ref{fig_sSIRF_with_without_convol_x500_yesdispersion} whose layout is identical to that of Fig.~\ref{fig_sSIRF_with_without_convol_x500_nodispersion}.
The first comments are on the transient of the real pusle ($\Delta t=$100 fs).
Dispersion in water has a weak influence on the real pulse signal in the two cases because the diameter of the scatterer ($\approx$\SI{100}{\micro\meter}) is much smaller than the distance over which the pulse spreads significantly (see Fig.~\ref{fig_isolines_ds_vs_L0_dt}).
This results suggests that at the scale of the scatterer, the pulse spreading could be neglected and hence, the SSIRF could be computed with a constant refractive index.
The influence of dispersion and extinction on the SSIRF is very strong for both the droplet  and the bubble, especially for $\Delta t_{v}\le \tau_0$ where the unconvolved (top) time signals show an unacceptable large background noise far above the transient signal of the 100 fs pulse.
Despite this very distorted signal for the SSIRF, when it is convolved with the pulse, it matches the transient of the real pulse very well (bottom).
Again, this very good agreement is explained by the fact that the very noisy time signal of the SSIRFs is due to the large variations of the refractive index over the unrealistic large range of $\lambda$. However the convolution with the pulse act as a frequency filtering to keep only the frequencies of the pulse, where the refractive index does not vary much, hence the good agreement.
This point also shows that even though the SSIRF computed with $\Delta t_v < \tau_0$ does not represent any physical phenomenon because of the zero-area rule, its convolution with the realistic signal of the pulse leads to a realistic results because the convolution filters out unphysical artifacts of the SSIRF.
Additional test were performed with $\Delta t_v = \tau_0/4$ and $\tau_0/8$ and the agreement was always very good.
Hence the SSIRF computed with $\Delta t_v < \tau_0$ can be considered as non-real object that needs to be convolved in order to carry a physical signal.
\begin{figure}[!htb]
	\centering
	\includegraphics[width=\columnwidth,keepaspectratio]{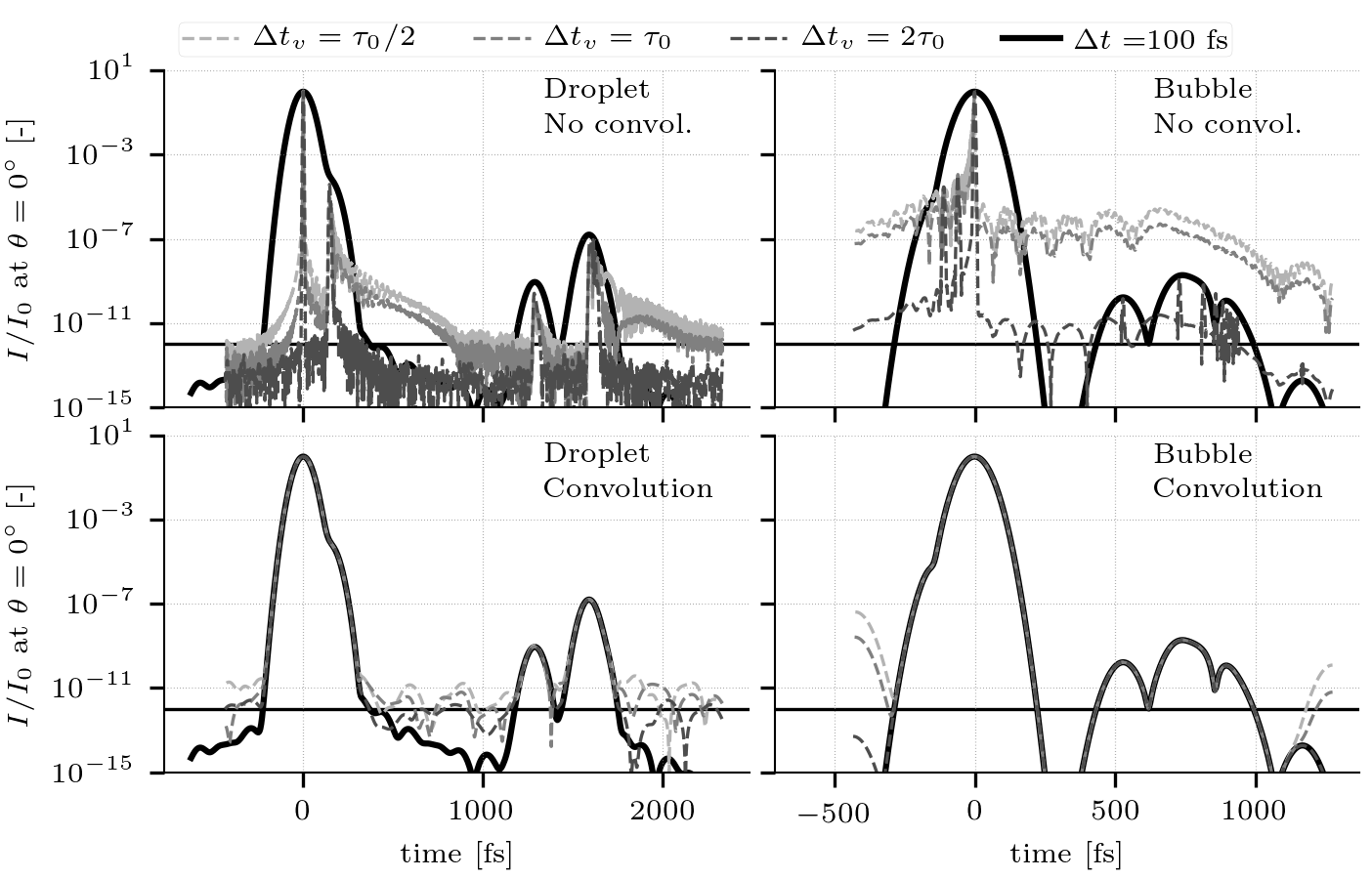}
	\caption{Same legend as Fig.~\ref{fig_sSIRF_with_without_convol_x500_nodispersion} except that water refractive index depends on the wavelength.}
	\label{fig_sSIRF_with_without_convol_x500_yesdispersion}
\end{figure}
\\\\
Finally, the SSIRF is illustrated in Fig.~\ref{fig_illustration_SSIRF_map} for a scatterer of \SI{100}{\micro\meter} for constant refractive indices and $\Delta t_v=2 \tau_0$.
The case of a water droplet in air (left) was resolved with a total resolution (including zero padding) $N_{\omega}$ of 2380337 while the case of air bubble in water (right) was resolved with $N_{\omega}=595085$. The better convergence of bubble is clearly visible. 
\\To conclude on this part, the SSIRF-based approach was shown to provide a very good transient time signal.

\section{Validation of transient single scattering with the energy approximation\label{sec_valide_energy_approx}}

In this part we verify that the SSIRF approach also works with the energy approximation, with the two methods discussed earlier. Also, we verify that the approaches work with a chirped pulse, \ie when the scatterer is located at a distance $L$ from the laser source. We only study air bubbles in water in this part. 
To validate the approaches, we use the intensity given by the exact transient scattered field of a chirped pulse after one scattering event (squared modulus of Eq.~\ref{eq_scatter_wave_disp}) as a reference, and we compare the methods expressed by their intensity time signal.
For the full Monte Carlo method (Method~1), it consists in randomly drawing a large number of scattering time (Eq.~\ref{eq_fullMC_time}). In the present case of single scattering, the time at which the scattered photon reaches the virtual detector (shown in Fig.~\ref{fig_sketch_geom}) is:
\begin{equation}
\left. T_I \right|_{(L,\Theta_0)} = \left. T_{g'} \right|_{L} + \left.  T_0 \right|_{\Theta_0}
\label{eq_singlescatt_MC_method}
\end{equation}
which is the summation of the random time due to the pulse chirped after a distance $L$ and the random time due to the scattering in the direction $\theta$. The intensity is given by the histogram of $T$.
The PDF of $\left. T_{g'} \right|_{L}$  is given by the intensity of the chirped pulse $I_p(t,L)=|g'(t,L)|^2$ where $g'(t,L)$ is computed from Eq.~\ref{eq_envelop_disp_new_frame}. The random variable $\left.  T_0 \right|_{\Theta_0}$ follows the PDF given by the energy of the SSIRF $| \bm{\phi}_S(t,\Theta_0)|^2$ (Eq.~\ref{eq_SSIRF_kernel}) with a duration $\Delta t_v$ of two cycles ($2\tau_0$).\\
For Method~2, the intensity is given by Eq.~\ref{eq_multiscat_energy_approx}, which in this case simplifies to:
\begin{equation}
I(t,\theta,L)  \approx I_p(t,L)  * I_{\phi}(t,\theta)
\label{eq_singlescatt_convol_meth}
\end{equation}
where $I_p(t,L)$ is the energy of the chirped pulse and $I_{\phi}(t,\theta)$ is the time signal of the SSIRF with the comb representation. We will show that detecting the peaks from the SSIRF computed with a constant index (non-dispersive medium) is preferable than in a dispersive medium. Indeed as shown earlier, the SSIRF computed with dispersion is usually noisier and more cumbersome to detect peaks. This approximation is also justified if the dimension of the bubbles are smaller than the characteristic length of the pulse spreading, \ie when the incident wave does not spread much between the two boundaries of the bubble. This result is important because it removes the necessity to characterize the refractive index over a large range of wavelengths.

\begin{figure}[!htb]
	\centering
	\includegraphics[width=\columnwidth,keepaspectratio]{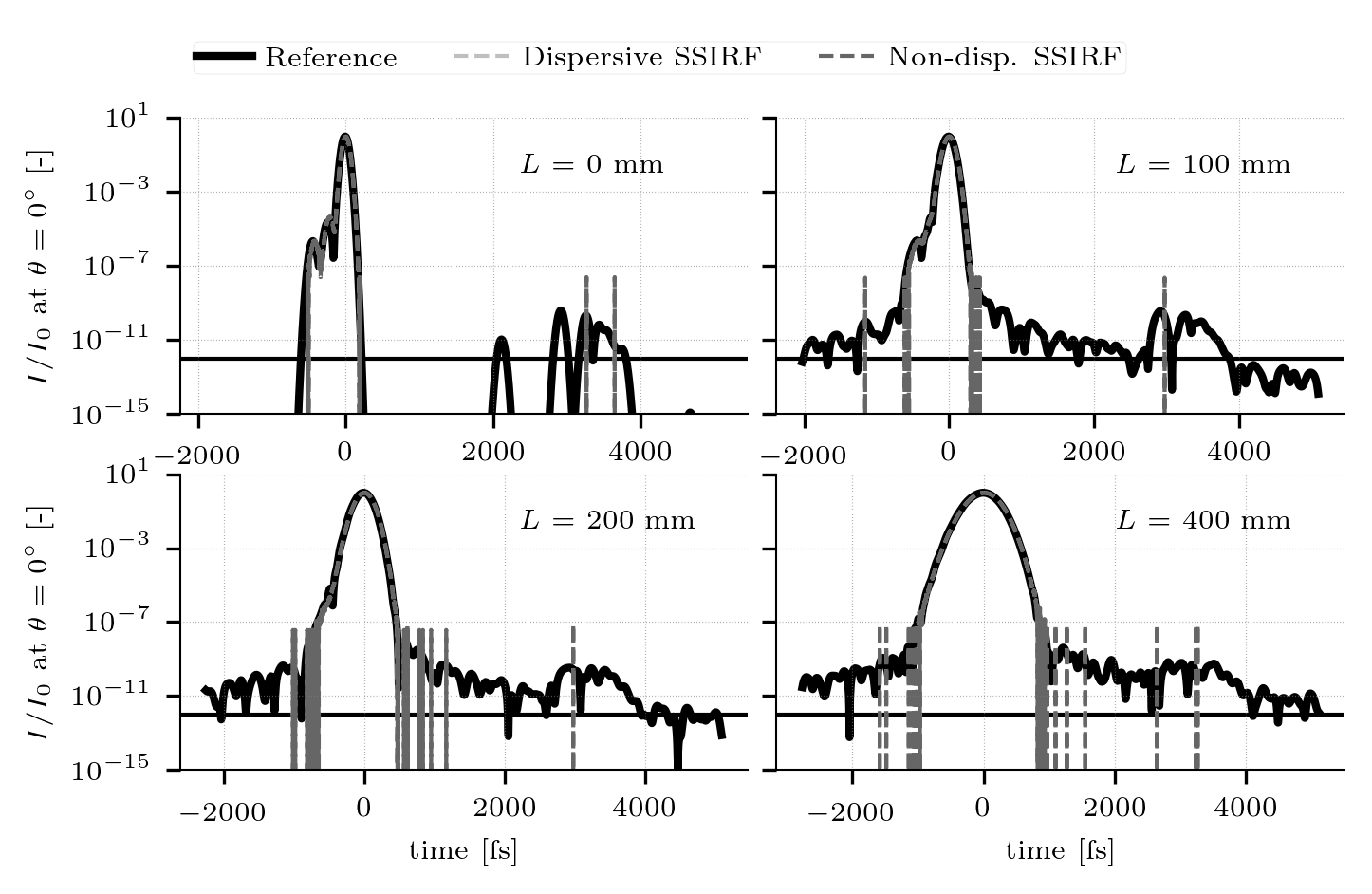}
	\caption{Time signal at $\theta=0^{\circ}$ for bubbles ($x=2000$) at different distances $L$ from the laser source with Method~1. The SSIRF was computed with varying (light grey curve) and constant (dark grey curve) dispersive index.}
	\label{fig_SSIRF_MonteCarlo_with_without_dispersion_x2000}
\end{figure}

\begin{figure}[!htb]
	\centering
	\includegraphics[width=\columnwidth,keepaspectratio]{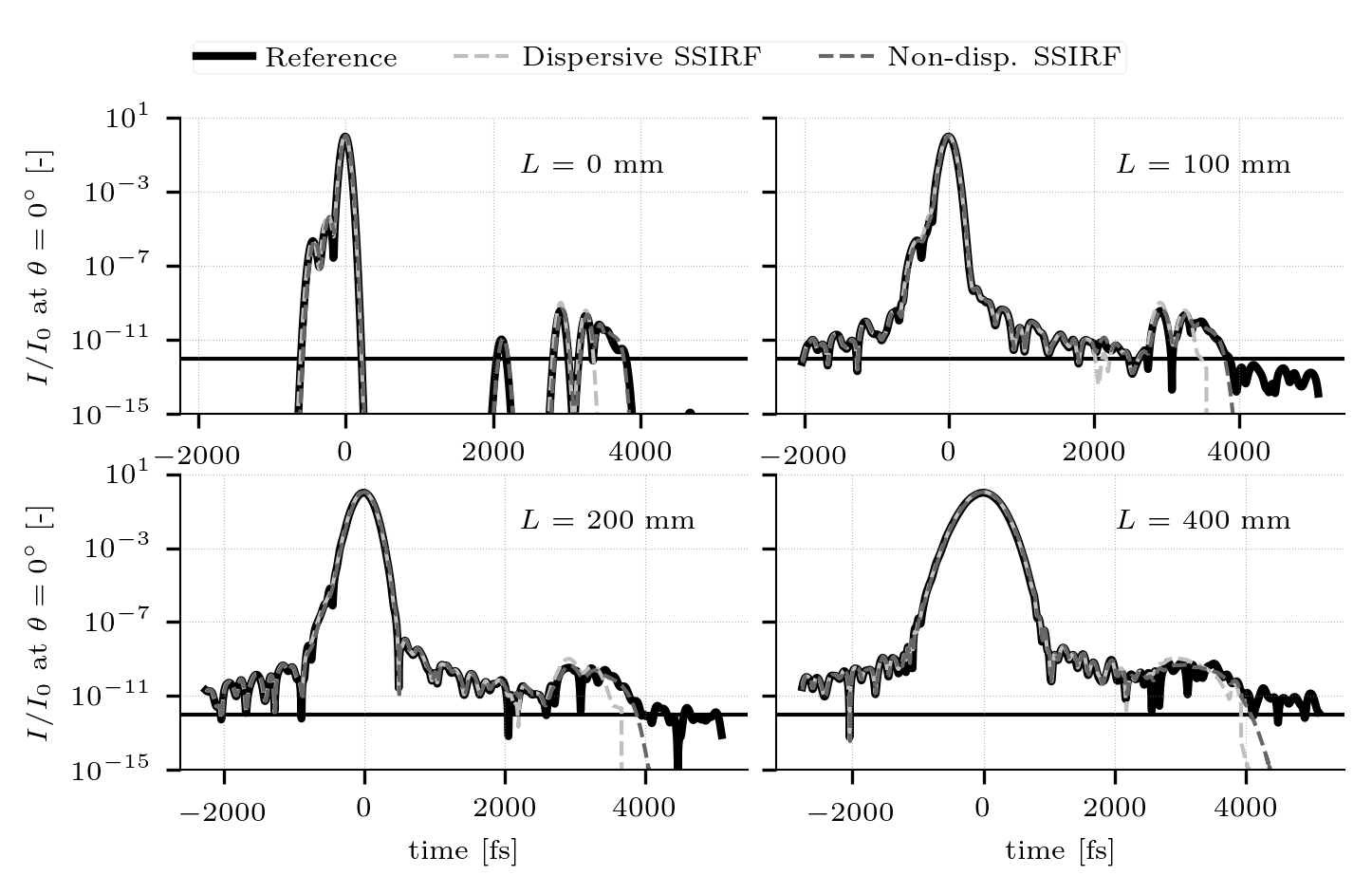}
	\caption{Time signal at $\theta=0^{\circ}$ for bubbles ($x=2000$) at different distances $L$ from the laser source with Method~2. The SSIRF was computed with varying (light grey curve) and constant (dark grey curve) dispersive index.}
	\label{fig_SSIRF_convol_with_without_dispersion_x2000}
\end{figure}

Figure~\ref{fig_SSIRF_MonteCarlo_with_without_dispersion_x2000} compares the time signals based on Eq.~\ref{eq_singlescatt_MC_method} with the SSIRF computed in dispersive and non-dispersive medium, for different $L$ in the forward ($\theta=0^{\circ}$) direction. Each curves is based on one billion samples.
The major peak around $t=0$ is well predicted for all $L$. The secondary peaks at $t<0$ are in acceptable agreement for $L=0$~mm but they are less and less resolved as $L$ increases.
Because of the very large amplitude ratio ($\approx$\dixmoins{9}) between majors and secondary peaks at $t>0$, several tens of billions samples would be necessary to reach statistical convergence for all peaks, which was not done here, and thus they are not captured. This is especially detrimental at $L=$~0 and \SI{100}{\milli\meter} where the secondary lobes are particularly visible.
\\Figure~\ref{fig_SSIRF_convol_with_without_dispersion_x2000} compares the time signals based on Eq.~\ref{eq_singlescatt_convol_meth} in the same condition as for Fig.~\ref{fig_SSIRF_MonteCarlo_with_without_dispersion_x2000}.
In this case, the agreement is very good for all distances. The secondary peaks at $t>0$ are well captured up to $L=$~\SI{200}{\milli\meter}, which is an advantage over Method~1.
However as for Method~1 the resolution of the secondary peaks at $t<0$ decreases as $L$ increases.
The results are slightly better with the non-dispersive SSIRF because more peaks are captured.

\section{Transient multiple scattering with the energy approximation\label{sec_transient_multi_energy}}

\subsection{Comparison with results from literature \label{ssec_valid_compare_calba}}

\begin{figure}[!htb]
	\centering
	\includegraphics[width=\columnwidth,keepaspectratio]{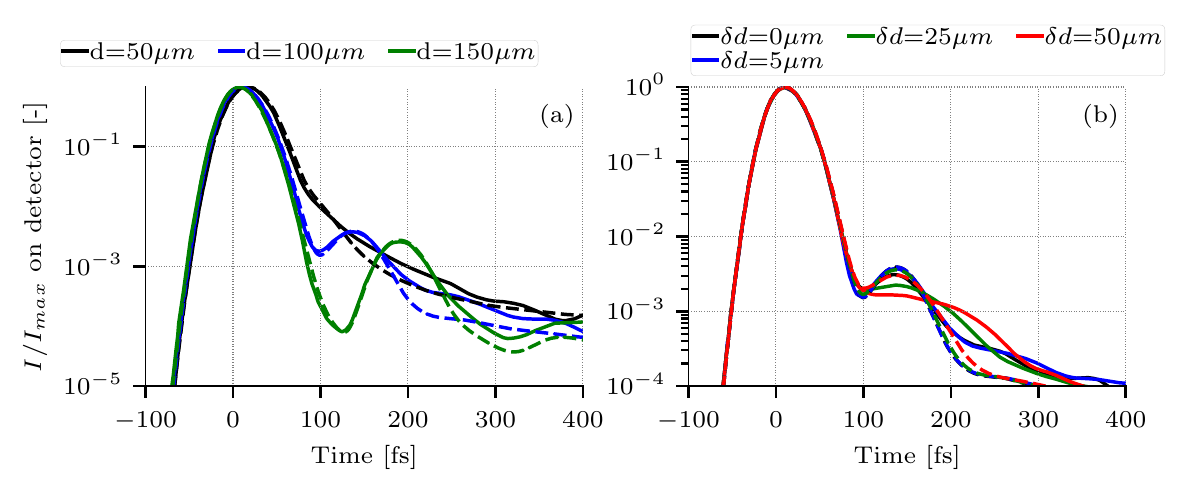}
	\includegraphics[width=\columnwidth,keepaspectratio]{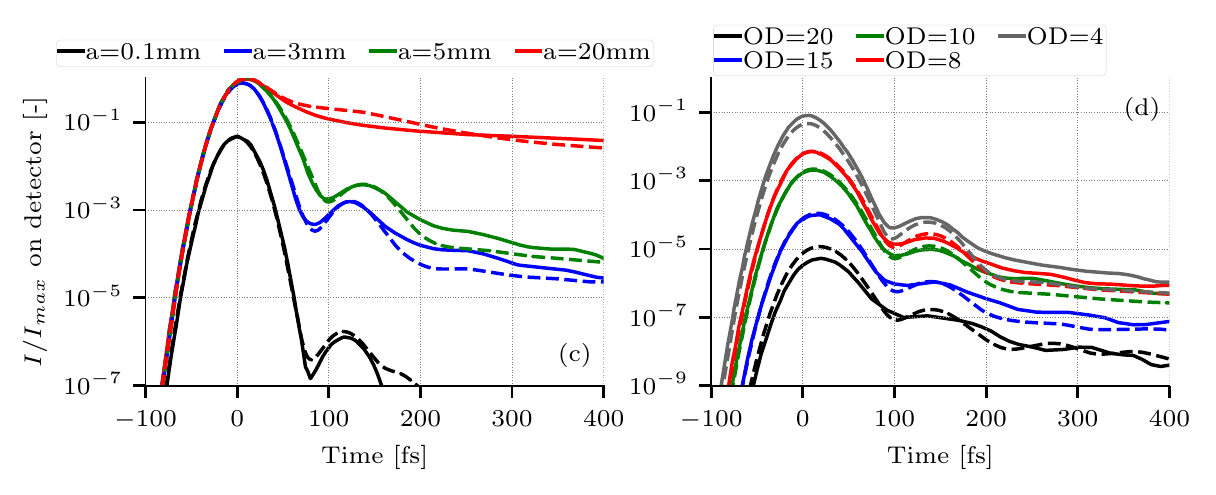}
	\caption{Intensity on the detector for different mean diameters (a), polydispersity (b), detector sizes (c) and optical thickness (d). Plain lines are from \cite{calba_monte_2006}, dashed lines are from the present Method~2. Reference droplets cloud is polydisperse with a Gaussian distribution of mean \SI{100}{\micro\meter} and width \SI{5}{\micro\meter}.}
	\label{fig_compare_calba}
\end{figure}

The results of Method~2 are compared with those of \cite{calba_monte_2006} where the authors used the energy approximation with the Comb Transport (our Method~2) to simulate the transient scattering of a polydisperse cloud of water droplets in air. The notable difference with the present approach is that they used the Debye expansion, and not the LMT, to create the comb time signal. There are several advantages to use the Debye expansion over the LMT. First, it requires less discretization points in the frequency domain to reach the same convergence for Eq.~\ref{eq_scatter_wave_disp}, provided that enough modes are computed ($\approx$ 20).
The authors report a number of point of $2^{11}$ for the Debye expansion versus $2^{18}$ for the LMT to reach the same accuracy.
Second, it is easier to detect peaks because each Debye mode is expressed as an individual time signal. Therefore different peaks of different refraction modes do not interfere or merge with each other.
Third, because each peak is associated to an order of refraction, it is possible to interpolate the amplitude and time of each refraction mode with regards to the angle of scattering and the droplet diameter. 
For instance if the set of randomly drawn diameter and angle $(d,\theta)$ is included in the square (in the parameter space) delimited by $(d_i,\theta_i)$ and $(d_{i+1},\theta_{i+1})$, then the amplitudes and delays of the peaks for each mode can be interpolated. With the LMT, because all modes are gathered in a single time signal, they cannot be easily identified, and hence the peaks cannot be interpolated.
The major drawbacks of the Debye expansion is that it does not converge in case of relative refractive index smaller than one (\eg air bubble in water), in the vicinity of the critical angle. This is the reason why the present study relies on LMT.
\\As in the present study, the strategy in \cite{calba_monte_2006} is to precompute the scattering phase functions that are used to randomly draw the scattering direction of photons.
The polar angle was discretized on a grid of $\mathrm d \theta=$~\SI{0.01}{\degree} resolution between 0 and \SI{1}{\degree}, and then $\mathrm d \theta=$~\SI{1}{\degree} up to \SI{180}{\degree}.
It was verified (but not presented here) that for scattering angles of 0 and 90$^{\circ}$ the time signal given by the Debye expansion in \cite{calba_monte_2006} matches the one of the LMT in the present study.\\
The configuration studied in \cite{calba_monte_2006} is a slab of thickness \SI{10}{\centi\meter} in the laser direction and of infinite extent in the other directions. The parameters of the reference case are as follows. The slab contains a polydisperse spray of water droplets in air, and is illuminated by a laser ($\lambda_0$~=~\SI{600}{\nano\meter}) pulse of FWHM \SI{50}{\femto\second}. The droplet size distribution is given by a Gaussian function of mean \SI{100}{\micro\meter} and width $\delta\!d$ at $e^{-1}$ of \SI{5}{\micro\meter}, and the optical depth is 8. A circular detector of diameter \SI{5}{\milli\meter} is located on the optical axis of the laser at \SI{20}{\centi\meter} from the slab.\\
Since Calba \etal \cite{calba_monte_2006} use the Comb Transport approach, we compare first their results with our Method~2 implemented in the code Scatter3D. All simulations in the present paper are performed with one billion photons.
The authors conducted several parametric studies which are used here for validation and presented in Fig.\ref{fig_compare_calba} where the intensity on the detector is normalized by its maximum and plotted versus time.

Plain and dashed lines correspond to the results of \cite{calba_monte_2006} and our results, respectively.
The time reference ($t=0$) is the one of ballistic photons reaching the detector.
The parameter studies were performed on the mean diameter of the spray (a), its polydispersity (b), the optical thickness (c) and the detector size (d).
The global shape of the curves is a smooth peak of highest amplitude followed by a second weaker one.
As shown in \cite{calba_monte_2006}, the first peak corresponds to ballistic (not scattered) and snake (only diffracted) photons, which explain why the first peak occurs slightly later than $t=$~\SI{0}{\femto\second}.
The second peak is due to photons that underwent one refraction and multiple diffractions.
These peaks are always well captured in time and amplitude by our method, thus proving that phenomena of ballistic, snake and refracted photons are accounted with the same accuracy as in \cite{calba_monte_2006}.
In addition, our parameter studies match those of \cite{calba_monte_2006} for the mean diameter, the optical thickness and the detector size, thus ensuring an equivalent treatment of these parameters. 
The discrepancy for the study on polydispersity is for now not clear. To rule out any mistakes from the present model, it is demonstrated in the Appendix~C
that the diameter distribution is correctly taken into account.

\subsection{Comparison of the two methods with droplets and bubbles\label{ssec_compare_two_approaches}}

\begin{figure}[!htb]
	\centering
	\includegraphics[width=0.495\columnwidth,keepaspectratio]{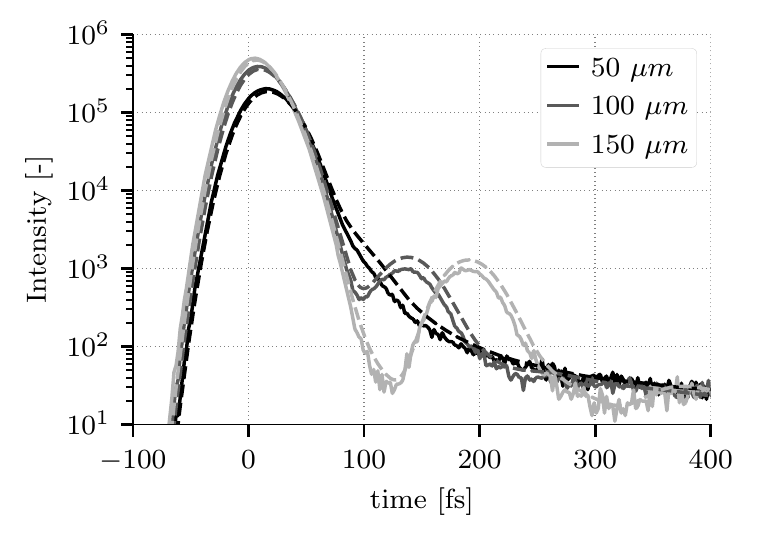}
	\includegraphics[width=0.495\columnwidth,keepaspectratio]{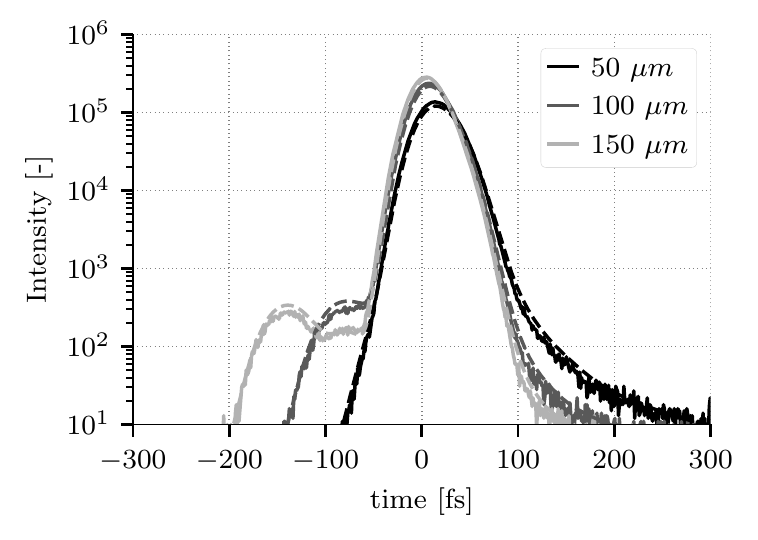}
	\caption{Comparison of the Monte Carlo approch (plain line) with the comb approach (dased line) for different mean droplet (left) and bubble (right) diameters.}
	\label{fig_compare_method_droplet_bubble}
\end{figure}

Method~1 and Method~2 presented here are compared to each other in the same configuration as the previous section, for different mean diameters. They are shown in Fig.~\ref{fig_compare_method_droplet_bubble} (left) where the intensity is in arbitrary units proportional to the detected number of photon. 
The first peak is larger for larger particles because of their larger cross section.
The prediction of the first peak is in agreement with only a slight shift in time of a few femtoseconds.
Except for the primary peak, the time signal of Method~1 is noisier than the one of Method~2, due to the lower statistical convergence inherent to Method~1.
Discrepancies are observed for the secondary peak with a shift in the amplitude, which means that in Method~1 the relative probability to have a refraction against having a diffraction in the forward direction is underestimated. Since both methods are based on the same computation of the SSIRF, the underestimation does not come from the SSIRF approach, but from the method itself. For larger time ($t>$~\SI{300}{\femto\second}), both methods match again, which suggests that the discrepancy lies in the way the method distribute in time the energy due to refraction.\\
The same simulation was performed while inverting liquid and gaseous phase, hence simulating a polydisperse cloud of bubbles in water. Dispersion was not taken into account. Results are shown in Fig.~\ref{fig_compare_method_droplet_bubble} (right).
First, we describe the differences with the water droplets case where the first peak is the result of ballistic and snake photons while the next peaks are due to refracted photons.
In the case of bubbles, the first peak to appear is the one attributed to photons which underwent only one refraction and any number of diffractions. This is because photons travel faster in air than in water, so that refracted photons travel faster than diffracted photon, as observed in single scattering simulation \cite{chaussonnet_scattering_2020}.
Other peaks earlier than \SI{-200}{\femto\second} were also observed (but not reported here) with an intensity $<10$. These are attributed to multiple refractions and any number of diffractions, which reduce further the time of flight of the photon.
The peak due to ballistic and snake photons only has still the largest amplitude and appears later. After that, the time signal contains no significant information.
Concerning the comparison of the two methods, the case of bubble allows to separate two effects.
First, the tail of the largest peak decreases faster with Method 1 compared to Method 2, which again suggests that Method~1 over promotes diffraction.
Second, the refraction peaks are predicted at the same time for the two methods while their amplitude is different, which confirm that Method~1 underestimates the relative probability of refraction in forward direction.
\\Concerning computational overheads, Method~2 is more expensive because of the discrete convolution on peaks for each photon whereas Method~1 only draw random times. However, since the precomputed table are discretized in time for Method~1 whereas only peak time location are saved in Method~2, Method~1 is extremely memory intensive compared to Method~2. For instance the results of Method~1 were obtained by discretizing the time over 5000 elements, resulting in precomputed tables of 46~MB versus 124~KB for Method~2. This large amount of memory can significantly slow the computation with Method~1, even more than with Method~2. With a case presented in this section, with the polydispersity discretized on 31 diameters (\ie 31 precomputed tables) and time discretized over 5000 elements, Method~1 was 13\% slower than Method~2.

\section{Scattering through a bubbly flow with pulse dispersion\label{sec_num_exp}}

In this section we illustrate the effect of the pulse spreading due to beam chirping in a more realistic configuration. The laser wavelength is set to \SI{800}{\nano\meter}, the slab thickness is \SI{100}{\milli\meter}. According to Fig.~\ref{fig_isolines_ds_vs_L0_dt}, we choose a pulse duration $\Delta t =$ \SI{100}{\femto\second} to minimize the pulse spread for a pathlength of \SI{100}{\milli\meter}. We assume the bubble size distribution is Gaussian with mean $d=$~\SI{500}{\micro\meter} and standard deviation $\delta\!d =$~\SI{100}{\micro\meter}. As in the previous section, we also investigate the influence of mean diameter and polydispersity on the time signal. The detector is as previously a disk of \SI{5}{\milli\meter} diameter located at \SI{200}{\milli\meter} from the slab.

\begin{figure}[!htb]
	\centering
	\includegraphics[width=\columnwidth,keepaspectratio]{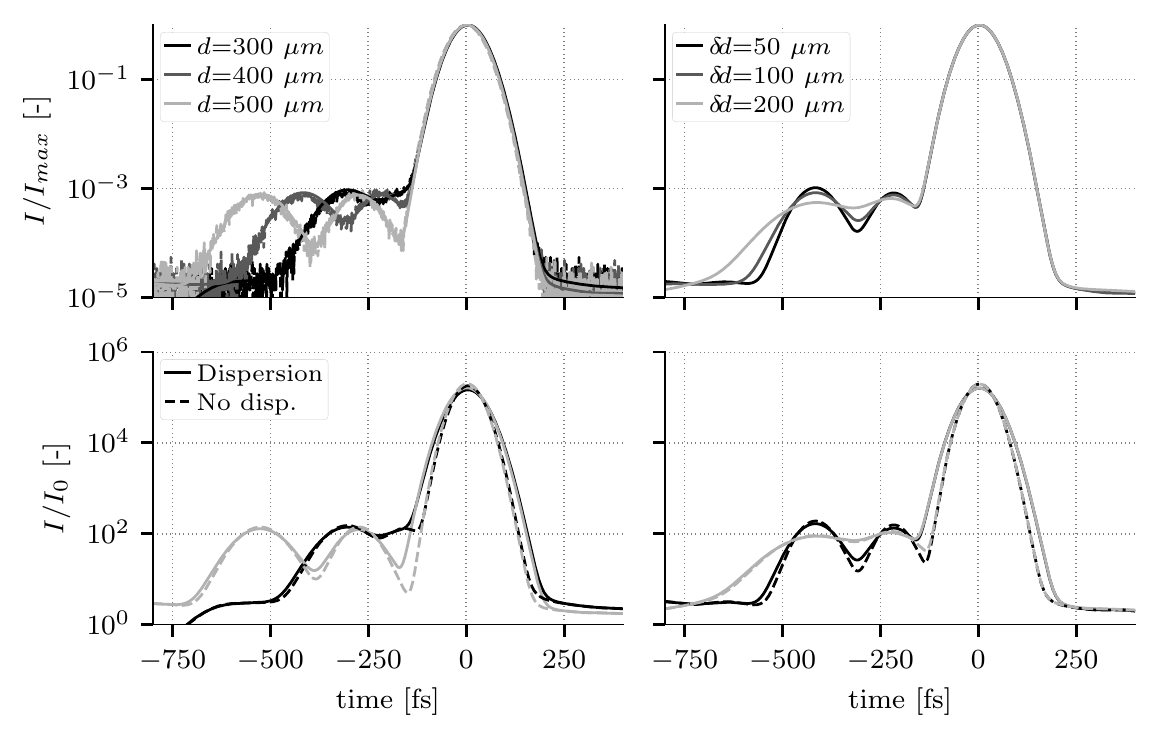}
	\caption{Top left: Comparison of the Monte Carlo approch (plain line) with the comb approach (dased line) for different mean droplet diameters. Top right: Comb approach for different polydispersity at mean diameter \SI{400}{\micro\meter}. Bottom: influence of dispersion for $d$ = 300 and \SI{500}{\micro\meter} (left) and $\delta d$ = 50 and \SI{200}{\micro\meter} (right)}
	\label{fig_real_bubbly_flow}
\end{figure}

To take the pulse spreading into account, we precompute the pulse spread $\Delta s(\Delta t,L)$ for a optical path length ranging from the minimum slab thickness 100 to \SI{500}{\milli\meter} which we load into Scatter3D. When the photon reaches the detector the optical path length is converted into the pulse spread.
For Method~1, the random variable of the pulse delay $ T_{g'} |_L$ (see Eq.~\ref{eq_random_var_meth1_multi}) is drawn according to a normal distribution whose variance is directly related to the spread.
For Method~2 we make a simplification. In principle, the total time signal on the detector is the summation of all photons, whose individual time signal is convolved by the chirped pulse. Hence, summing Eq.~\ref{eq_multiscat_energy_approx} on all photons reaching the detector writes:
\begin{equation}
I_{\text{detector}}(t)  =
\sum_{\text{photons}} \left(
I_p(t,L)  * \bigotimes_i  I_{\phi,d_i}(t,\theta_i) \right)
\label{eq_multiscat_energy_approx2}
\end{equation}
In the present configuration, since the angle resolution of the detector is rather small ($\Delta \theta \approx 0.712^{\circ}$), photons of similar time of arrival on the detector have a similar optical path length $L$ and hence a similar pulse spread $\Delta s$.
In other words the photons that reach the detector at time $t$ belong to a pulse of mean spread $\Delta s_m(t)$. This allows us to take the convolution by $I_p(t,L)$ out of the summation in Eq.~\ref{eq_multiscat_energy_approx2}.
Therefore, when the photon reach the detector its time signal is not convolved by the chirped pulse, but directly summed up with other photons. Also, we record the mean pulse spread $\Delta s_m(t)$ for each $t$, weighted by the peaks amplitude.
At the end of the simulation, the time signal is a very dense comb, that we convolve with the locally chirped pulse:
\begin{equation}
\begin{split}
I_{\text{detector}}(t,\theta,L)  & \ \approx 
\int_{t_{\min}}^{t_{\max}}
S(\tau)
\cdot 
I'_p(t-\tau, \Delta s_m(t)) \, \mathrm d\tau \\
\text{where} \quad
S(\tau) = & \  \left( \sum_{\text{photons}} \bigotimes_i  I_{\phi,d_i}(\tau,\theta_i) \right)
\label{eq_mean_pulse_spread}
\end{split}
\end{equation}
and $I'_p(\tau,\Delta s_m(t))$ is the chirped pulse intensity of spread $\Delta s_m(t)$ at time of arrival $t$. Note that $S(\tau) $ is the dense comb obtained at the end of the simulation.
\\\\
The parameter studies on the mean diameters is shown in Fig.~\ref{fig_real_bubbly_flow} (top left) with the two methods where their curves of same color match very well.
As in the previous section, ballistic and snake photons carry most of the energy and reach the detector after the refraction photons. 
Two peaks for $d=$~400 and \SI{500}{\micro\meter} are observed prior to the no-refraction mode. The earliest one corresponds to $p=1$ mode of the Debye expansion while the second corresponds to other $p>2$ modes as explained in \cite{chaussonnet_scattering_2020}. These two peaks in forward direction are also visible in the illustration of the SSIRF for bubbles (Fig.~\ref{fig_illustration_SSIRF_map} right).
The influence of dispersion is illustrated in Fig.~\ref{fig_real_bubbly_flow} (bottom left) where extreme cases ($d=$~300 and \SI{500}{\micro\meter}) are shown with and without dispersion.
The influence of the pulse spread is stronger for thinner peaks (\eg main peak) and almost negligible for wider peaks (\eg  $p=1$ peak). 
For $d=$~\SI{300}{\micro\meter} the $p>2$ peak is merged to the main peak.
Therefore, when the pulse spread increases (\eg for a thicker slab) the different refraction modes could be less distinguishable.
The parameter study on the polydispersity is shown in Fig.~\ref{fig_real_bubbly_flow} (top right) for Method~2 only. The general trend is that wider diameter distributions lead to smoother time signals, thus diminishing the contrast of the refraction modes. 
The same effects due the pulse spread is observed.
Note that in our representation the influence of extinction is neglected.\\

To conclude this part, the mean diameter of the bubble cloud influences the temporal position of the peaks while the width of the diameter distribution acts on the width of the refraction modes. 
As for the previous results on droplets claimed by \cite{calba_monte_2006},
the characterization of the bubble cloud could be obtained from the time signal, with an appropriated analysis depending on the exact configuration, and on unknown parameters. Note that concentration and optical depth also influence the peaks amplitudes and their width.

\section{Conclusion}

In this paper a model of the transient of multiple scattering in a dispersive medium was presented in terms of EM wave amplitude and turned to intensity.
The model is more generic than the current state of the art because it relies on the Lorenz-Mie Theory and hence allows to consider clouds of scatterers of relative refractive index lower than one, typically air bubbles in water.
It can be further extended to take polarization into account.\\
We showed that the scattering effect can be decoupled from the pulse generation and propagation and hence it can be modeled individually by the Scattering Impulse Response Function (SIRF). The SIRF can be accurately approximated by a smoothed expression (SSIRF) to detect the peaks in the time-direction map.
The SSIRF was turned into an energy form to be incorporated in the Radiative Transport Equation.
The maps of the pulse spread and extinction were drawn for ultra short pulse propagating in water, and an analytical expression for the pulse spread was given and validated for visible light.
Two methods to account for the transient were investigated in the framework of Monte Carlo simulation. The first being a naive Monte Carlo approach, where scattering delays are randomly drawn, the second being the transport of the time signal of the scattering transient. It was found that even slightly more complex to program, Method~2 is much more efficient in terms of resolution, statistical convergence, time of execution and memory consumption. Therefore the authors strongly advise the use of Method~2.
The two methods were validated against previous numerical simulations from the literature.
When applied to the case of multiple scattering by a cloud of bubbles in water, it was shown that the scattered photons exit the medium earlier than ballistic and snake photons, and even that 
different peaks occur for photons undergoing a different number of refraction. These results open the door for new diagnostics based on ultrashort laser pulse to characterize bubbly flows.
 
\section*{Funding}
The authors acknowledge the support of the US Office of Naval Research (N000141712616) under the supervision of Drs. Thomas Fu and Woei-Min Lin, and the US Department of Energy (DE-NE0008747).

\appendix

\section*{Appendix A: Details on pulse spreading in water\label{appendix_pulsespread}}
\begin{figure}[!htb]
	\centering
	\includegraphics[width=\columnwidth,keepaspectratio]{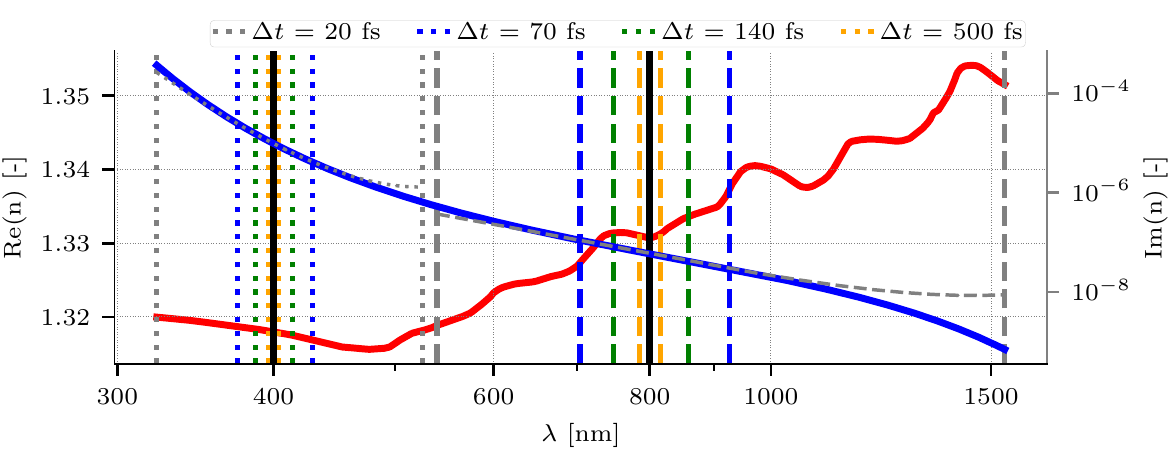}
	\caption{Real (thick plain blue line) \cite{harvey_revised_1998} and imaginary (thick plain red line) \cite{segelstein_complex_1981} part of water refractive versus the wavelength. Vertical lines show the spectrum boundaries for various duration at $\lambda_0 =$~400 (dotted lines) and \SI{800}{\nano\meter} (dashed lines). Grey dotted, and dashed non vertical lines are second order Taylor expansion of the real part, centered at $\lambda_0 =$~400 and \SI{800}{\nano\meter}.}
	\label{fig_n_water_versus_lambda_alloverlaid}
\end{figure}

We plot the real (thick plain blue line) \cite{harvey_revised_1998} and the imaginary (thick plain red line) \cite{segelstein_complex_1981} part of water refractive versus the wavelength in Fig.~\ref{fig_n_water_versus_lambda_alloverlaid}. We overlay the spectrum boundaries (vertical lines) of a light pulse for various duration $\Delta t$, with a central wavelength at 400 (dotted lines) and \SI{800}{\nano\meter} (dashed lines). These wavelengths correspond to a doubled and single frequency Ti:Sapphire laser, respectively.
The grey dotted, and dashed non vertical lines are second order Taylor expansion of the real part, they are discussed later.\\
First, the wavelength boundaries are not equally centered around $\lambda_0= 2 \pi / \omega_0$ because the wavelength is inversely proportional to $\omega$: $\lambda = 2 \pi c / (\omega_0 \pm \Delta \omega)$, so that the small frequencies $\omega_{\min}$ dramatically increase $\lambda_{\max}$, as particularly visible with $\Delta t$~=~\SI{20}{\femto\second}.
Shorter pulses lead to larger wavelength ranges, and hence to larger dispersion which eventually lead to larger pulse spreading.
Concerning the extinction for $\lambda_0=$\SI{800}{\nano\meter}, it increases by more than two orders of magnitude from 1000 to \SI{1400}{\nano\meter}, so that the intensity of pulses below \SI{70}{\femto\second} are more damped as visible in Fig.~\ref{fig_isolines_ds_vs_L0_dt}.
In the case of $\lambda_0=$\SI{400}{\nano\meter}, the extinction is almost two orders of magnitude smaller for all investigated pulse duration compared to the extinction at \SI{800}{\nano\meter}, hence the larger propagated intensity (Fig.~\ref{fig_isolines_ds_vs_L0_dt}).
\\\\
In order to demonstrate that the optimal $\Delta t$ to minimize $\Delta s$ is proportional to $\sqrt{L}$, we express the dispersion not as $n(\lambda)$, but as $k(\omega)$, $k$ being the wavenumber. This is equivalent, and more appropriate for wave propagation.
A Taylor expansion of $k$ at $\omega_0$ to the second order gives:
\begin{equation}
k(\omega) \approx k_0 + k'_0 (\omega - \omega_0) + \frac{1}{2} k''_0 (\omega - \omega_0)^2
\label{eq_quadratization_komega}
\end{equation}
where
\begin{equation}
k_0 = k(\omega_0)
\ \text{,} \quad
k'_0 = \left.\frac{\mathrm d k}{\mathrm d \omega}\right|_{\omega_0}
\quad \text{and} \quad
k''_0 = \left.\frac{\mathrm d^2 k}{\mathrm d \omega^2}\right|_{\omega_0}
\label{eq_all_k_of_omega}
\end{equation}
Note that Eq.~\ref{eq_quadratization_komega} can be expressed in term of $n(\lambda)$ by: %
\begin{equation}
k_0 = \frac{2 \pi} {\lambda_0} n_0
\ \text{,} \quad
k'_0 = \frac{n_0}{c_0} \left(1 - \lambda_0 \frac{n'_0}{n_0} \right)
\quad \text{and} \quad
k''_0 = \frac{\lambda_0^3}{2 \pi c_0^2} n''_0
\label{eq_all_k_of_lambda}
\end{equation}
where $n_0$, $n'_0$ and $n''_0$ have the same definition as in Eq.~\ref{eq_all_k_of_omega} by substituting $(k,\omega)$ with $(n,\lambda)$:
\begin{equation}
n_0 = n(\lambda_0)
\ \text{,} \quad
n'_0 = \left.\frac{\mathrm d n}{\mathrm d \lambda}\right|_{\lambda_0}
\quad \text{and} \quad
n''_0 = \left.\frac{\mathrm d^2 n}{\mathrm d \lambda^2}\right|_{\lambda_0}
\label{eq_all_n_of_lambda}
\end{equation}
One can show that if the original non-chirped pulse is expressed in the time domain as $\e^{-(t/\gamma_0)^2}$, the actual time constant $\gamma(z)$, representative of the spread in time of the pulse at coordinate $z$ is expressed as \cite{orfanidis_electromagnetic_2016}:
\begin{equation}
\gamma(z)^2 = \gamma_0^2 + \left( \frac{2 k_0'' z}{\gamma_0} \right)^2
\label{eq_all_n_of_lambda2}
\end{equation}
As the spread in space of the pulse is proportional to the spread in time, minimizing the spatial spread is equivalent to minimize the time spread. Therefore, solving
$\partial \gamma / \partial \gamma_0= 0$ leads to $\gamma_0 = \sqrt{2 k_0'' z}$. In terms of pulse duration $\Delta t_{\min\text{@}L}$ that minimizes the spread at a given location $L$:
\begin{equation}
\Delta t_{\min\text{@}L} = 2 \sqrt{2 \log (2) k_0'' L}
\label{eq_all_n_of_lambda3}
\end{equation}
Expressing $\Delta t_{\min\text{@}L} = K \sqrt{L}$ in femtoseconds and millimeters leads to a constant $K$ of 21.91 and 11.58 for a wavelength of 400 and \SI{800}{\nano\meter}, respectively. This corroborates the correlation of Fig.~\ref{fig_isolines_ds_vs_L0_dt}, and confirms that in the investigated variation ranges, the real part of the water refractive index can be approximated by its second-order Taylor expansion in $\lambda$, as shown in Fig.~\ref{fig_n_water_versus_lambda_alloverlaid}.
More generally, we estimate the spatial spread of the pulse from the quadratization of the refractive index (Eq.~\ref{eq_quadratization_komega}) as:
\begin{equation}
\Delta s(\Delta t_0,z) = \frac{c_0}{n_0} \sqrt{\Delta t_0^2 + [8\log(2) k''_0 z / \Delta t_0]^2}
\label{eq_ds_from_quadra}
\end{equation}
whose deviation from Eq.~\ref{eq_incident_wave_disp_new_frame} is relatively constant (between 2.83 and 2.90\%) for $\lambda_0$~=~\SI{400}{\nano\meter}.
For $\lambda_0$~=~\SI{800}{\nano\meter}, the deviation is much heterogeneous, but not depicted here. For a pulse duration above \SI{100}{\femto\second}, the maximum error is ~2.5\% whereas it increases to 10\% when $\Delta t$ goes to \SI{20}{\femto\second}. These deviations could be considered as acceptable to use Eq.~\ref{eq_ds_from_quadra} to model the spread of a light pulse in the present conditions.

\section*{Appendix B: Details of the hybrid model for the real part of water refractive index\label{appendix_hybrid}}

The matching of the database of from Harvey \etal \cite{harvey_revised_1998} and Segelstein \cite{segelstein_complex_1981}
is achieved by a third order spline that matches the zeroth and first derivative of both models:
\begin{equation}
S(\lambda) = -7.724559 \times 10^{19} \lambda^3 + 1.697097 \times 10^{14}\lambda^2  - 6.023895 \times 10^{7} \lambda + 7.301508
\label{eq_hybrid_model}
\end{equation}
with $\lambda$ in meter.
The model is recalled in Table.~\ref{table_segel_harvey}
We consider that the refractive index of water does not vary for wavelengths larger than \SI{1}{\centi\meter}.
\begin{table}[!htb]
	\centering
	\caption{Hybrid model for the refractive index of water}
	\label{table_segel_harvey}
	\begin{tabular}{  l | c  c  c   c }
	$\lambda_0$ & 100 $-$ \SI{175}{\nano\meter}  & 175 $-$ \SI{200}{\nano\meter} & \SI{200}{\nano\meter} $-$ \SI{2.5}{\micro\meter} & \SI{2.5}{\micro\meter} $-$ \SI{1}{\centi\meter}\\
	\hline
	Database &  Segelstein \cite{segelstein_complex_1981} & Eq.~\ref{eq_hybrid_model} & Harvey \etal \cite{harvey_revised_1998} & Segelstein \cite{segelstein_complex_1981} \\
	\end{tabular}
\end{table}

\section*{Appendix C: Validation of polydispersity on simple cases\label{appendix_polydisp}}

\begin{figure}[!htb]
	\centering
	\includegraphics[width=0.495\columnwidth,keepaspectratio]{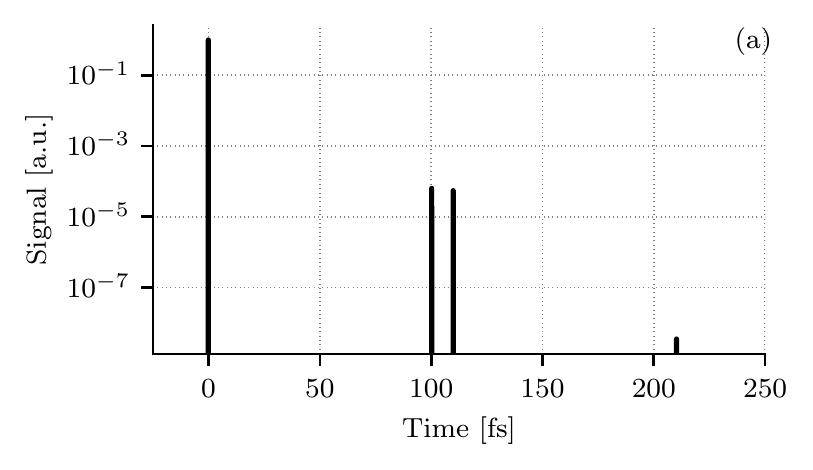}
	\includegraphics[width=0.495\columnwidth,keepaspectratio]{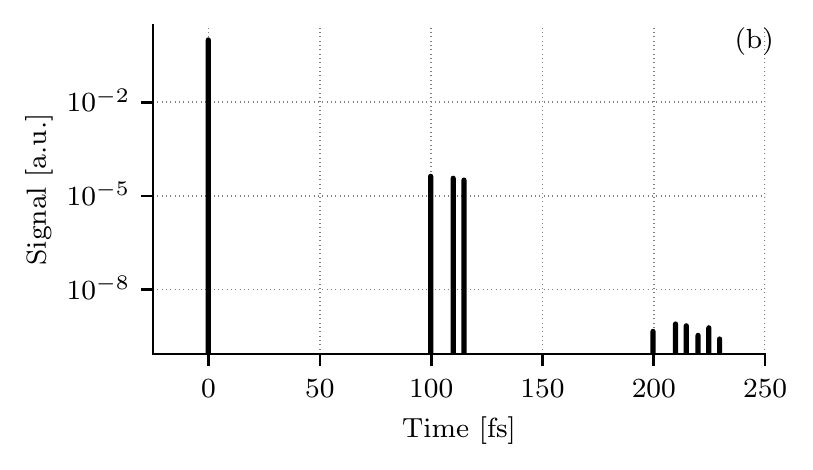}
	\includegraphics[width=0.495\columnwidth,keepaspectratio]{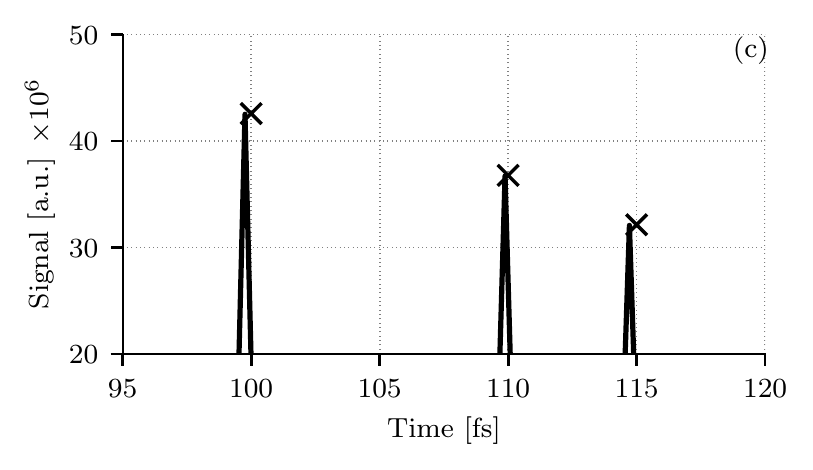}
	\includegraphics[width=0.495\columnwidth,keepaspectratio]{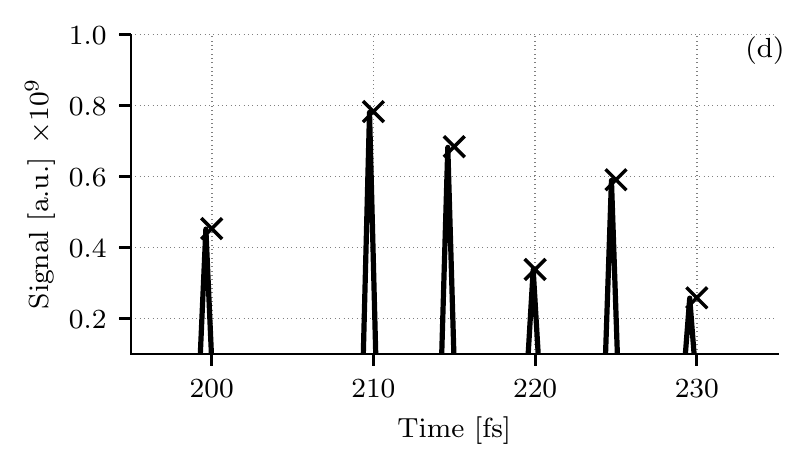}
	\caption{a) Example of a single photon. b) Signal for a uniform diameter distribution. c) Closeup for one-refraction peaks. d) Closeup for two-refraction peaks.}
	\label{fig_illustration_validation_polydisp}
\end{figure}

\begin{figure}[!htb]
	\centering
	\includegraphics[width=0.495\columnwidth,keepaspectratio]{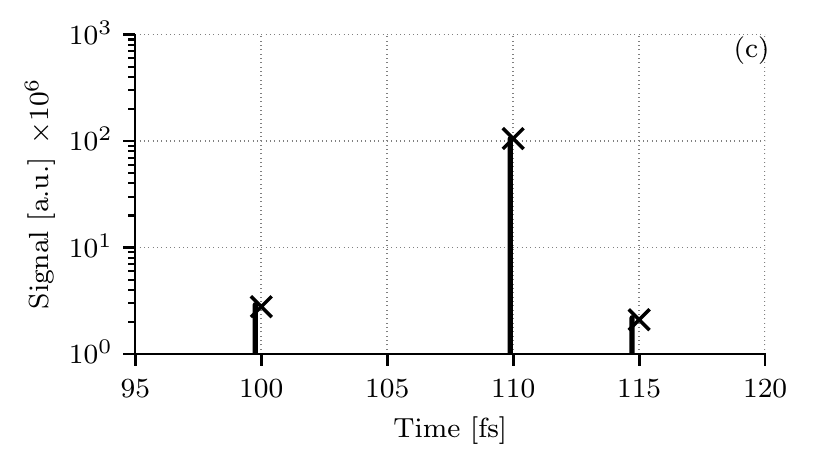}
	\includegraphics[width=0.495\columnwidth,keepaspectratio]{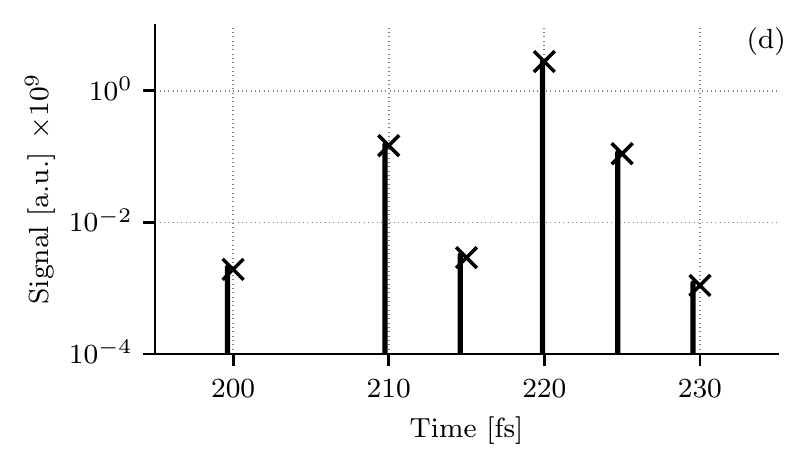}
	\caption{Left: closeup for one-refraction peaks. Right: closeup for two-refraction peaks.}
	\label{fig_illustration_validation_polydisp_gauss}
\end{figure}

We choose a simple case made of three different diameters, where each droplet scatters light only in forward direction ($\theta=0^{\circ}$) with two peaks, one for diffraction and one for refraction. Their time and normalized intensity are labeled $(t_{ij},I_{ij})$ for the diameter $d_i$ and $j$\textsuperscript{th} peak. Their numeric values are summarized in Table~\ref{tab_peaks}. The times of the second peak were slightly modified to ease the visualization.
\begin{table}[!htb]
	\centering
	\caption{Time (in femtosecond) and normalized intensity of the two peaks for each diameter.}
	\label{tab_peaks}
	\begin{tabular}{  l | c c c }
	& $d_0$ & $d_1$ & $d_2$ \\
	\hline
	Diffraction peak & (0, 1) & (0, 1) & (0, 1) \\
	Refraction peak & (100, \num{6.387e-5}) & (110, \num{5.517e-5}) & (115, \num{4.821e-5}) \\
	\end{tabular}
\end{table}
We limited the simulation to two scattering events exactly, thus leading to only one discrete convolution for the time signal. 
To post-process the result we did not convolve by the pulse signal in order to separate each peak.
Each numerical photon depicts a comb signal which is normalized so that its time integral is equal to an elementary amount of energy set arbitrarily to one. Since the time step is constant, in our simplified case we have
$(I_{i_00} + I_{i_01})(I_{i_10} + I_{i_11}) = 1$ for two scattering events.
As illustrated in Fig.~\ref{fig_illustration_validation_polydisp}(a), there are four peaks, one for pure diffraction, two for one refraction and one for two refractions. Their time is $0$, $t_{i_01}$, $t_{i_11}$ and $t_{i_01} + t_{i_11}$, respectively, and their probability is given by the previous equation.
The total signal is made of one pure diffraction peak of amplitude unity, a first group of three peaks due to one refraction and a second group of six peaks due to two refractions. An example is given in Fig.~\ref{fig_illustration_validation_polydisp} (b) for a uniform distribution.
The sum of all peaks is given by:
\begin{equation}
S = 1+ \sum_{(i,j)} p_i \, p_j (I_{d_i,1} + I_{d_j,1}) + \sum_{(i,j)} (p_i \, I_{d_i,1}) \, (p_j \, I_{d_j,1})
\end{equation}
where $p_i$ is the probability to have the diameter $d_i$. The peaks amplitude, depending on its group, is given in Table~\ref{tab_values_polydisp}.
First, we investigated the time signal when the three diameters are uniformly distributed, hence $p_i=1/3$ for all diameters. The time signal is shown in Fig.~\ref{fig_illustration_validation_polydisp}(b) where the peaks are due to one refraction are between $t=$~100 and \SI{120}{\femto\second} and those of two refraction are between $t=$~200 and \SI{235}{\femto\second}.
Their amplitude is estimated from Table~\ref{tab_values_polydisp} and match well the simulation as shown in Fig.~\ref{fig_illustration_validation_polydisp} (c) and (d).
\begin{table}[!htb]
	\centering
	\caption{Formulas to calculate times and amplitudes of the peaks due to refraction}%
	\label{tab_values_polydisp}
	\begin{tabular}{  l c c c }
	& Time & Amplitude & Indices \\
	\hline
	One diffraction group & $t_i $ & $ 2 p_i \, I_{d_i,1} / S$ & $i \in \llbracket 0,2 \rrbracket $ \\
	Two diffraction group (same diam.) &  $2 t_i$ & $ p_i^2 \, I_{d_i,1}^2 / S$  & $ i \in \llbracket 0,2 \rrbracket$ \\
	Two diffraction group (diff. diam.) &  $t_i + t_j$ & $2 \, p_i p_j \, I_{d_i,1} I_{d_j,1} / S$  & $(i,j)^2 \in \llbracket 0,2 \rrbracket^2, i \neq j$ \\
	\end{tabular}
\end{table}
The second case is made of three diameters following a Gaussian distribution. The two extreme diameters are at $3 \sigma$ while the middle one is on the mean. The probabilities are thus $p_0 = p_2 = 0.0228$ and $p_1 = 0.9545$. The amplitudes of the peaks are once again calculated with Table~\ref{tab_values_polydisp} and the good agreement is shown in Fig~\ref{fig_illustration_validation_polydisp_gauss}.
The results presented here ensure that the diameter distribution is correctly predicted by our model.

\bibliography{LorenzMieScattering}

\end{document}

%% file: sketch_of_configuration.pdf_tex
\begingroup%
  \makeatletter%
  \providecommand\color[2][]{%
    \errmessage{(Inkscape) Color is used for the text in Inkscape, but the package 'color.sty' is not loaded}%
    \renewcommand\color[2][]{}%
  }%
  \providecommand\transparent[1]{%
    \errmessage{(Inkscape) Transparency is used (non-zero) for the text in Inkscape, but the package 'transparent.sty' is not loaded}%
    \renewcommand\transparent[1]{}%
  }%
  \providecommand\rotatebox[2]{#2}%
  \newcommand*\fsize{\dimexpr\f@size pt\relax}%
  \newcommand*\lineheight[1]{\fontsize{\fsize}{#1\fsize}\selectfont}%
  \ifx\svgwidth\undefined%
    \setlength{\unitlength}{545.78881742bp}%
    \ifx\svgscale\undefined%
      \relax%
    \else%
      \setlength{\unitlength}{\unitlength * \real{\svgscale}}%
    \fi%
  \else%
    \setlength{\unitlength}{\svgwidth}%
  \fi%
  \global\let\svgwidth\undefined%
  \global\let\svgscale\undefined%
  \makeatother%
  \begin{picture}(1,0.33346174)%
    \lineheight{1}%
    \setlength\tabcolsep{0pt}%
    \put(0,0){\includegraphics[width=\unitlength,page=1]{sketch_of_configuration.pdf}}%
    \put(0.06636357,0.17551091){\color[rgb]{0,0,0}\makebox(0,0)[lt]{\begin{minipage}{0.08883899\unitlength}\raggedright Laser\end{minipage}}}%
    \put(0.19642937,0.31239876){\color[rgb]{0,0,0}\makebox(0,0)[lt]{\begin{minipage}{0.26235263\unitlength}\raggedright Propagation medium\end{minipage}}}%
    \put(0.44532516,0.06458654){\color[rgb]{0,0,0}\makebox(0,0)[lt]{\begin{minipage}{0.2387548\unitlength}\raggedright Reference path\end{minipage}}}%
    \put(0.69000758,0.24583546){\color[rgb]{0,0,0}\makebox(0,0)[lt]{\begin{minipage}{0.12215364\unitlength}\raggedright Particle\end{minipage}}}%
    \put(0,0){\includegraphics[width=\unitlength,page=2]{sketch_of_configuration.pdf}}%
    \put(0.52011131,0.19599207){\color[rgb]{0,0,0}\makebox(0,0)[lt]{\begin{minipage}{0.12215364\unitlength}\raggedright $L$\end{minipage}}}%
    \put(0.34485496,0.26169806){\color[rgb]{0,0,0}\makebox(0,0)[lt]{\begin{minipage}{0.19225318\unitlength}\raggedright Gaussian pulse\\ (plane wave)\end{minipage}}}%
    \put(0,0){\includegraphics[width=\unitlength,page=3]{sketch_of_configuration.pdf}}%
    \put(0.84780715,0.31796533){\color[rgb]{0,0,0}\makebox(0,0)[lt]{\begin{minipage}{0.12701205\unitlength}\raggedright Position of the virtual detector\end{minipage}}}%
    \put(0,0){\includegraphics[width=\unitlength,page=4]{sketch_of_configuration.pdf}}%
    \put(0.814366,0.20062126){\color[rgb]{0,0,0}\makebox(0,0)[lt]{\begin{minipage}{0.12215364\unitlength}\raggedright $\theta$\end{minipage}}}%
    \put(0,0){\includegraphics[width=\unitlength,page=5]{sketch_of_configuration.pdf}}%
  \end{picture}%
\endgroup%

%% file: sketch_MonteCarlo_Method1.pdf_tex
\begingroup%
  \makeatletter%
  \providecommand\color[2][]{%
    \errmessage{(Inkscape) Color is used for the text in Inkscape, but the package 'color.sty' is not loaded}%
    \renewcommand\color[2][]{}%
  }%
  \providecommand\transparent[1]{%
    \errmessage{(Inkscape) Transparency is used (non-zero) for the text in Inkscape, but the package 'transparent.sty' is not loaded}%
    \renewcommand\transparent[1]{}%
  }%
  \providecommand\rotatebox[2]{#2}%
  \newcommand*\fsize{\dimexpr\f@size pt\relax}%
  \newcommand*\lineheight[1]{\fontsize{\fsize}{#1\fsize}\selectfont}%
  \ifx\svgwidth\undefined%
    \setlength{\unitlength}{711.93644175bp}%
    \ifx\svgscale\undefined%
      \relax%
    \else%
      \setlength{\unitlength}{\unitlength * \real{\svgscale}}%
    \fi%
  \else%
    \setlength{\unitlength}{\svgwidth}%
  \fi%
  \global\let\svgwidth\undefined%
  \global\let\svgscale\undefined%
  \makeatother%
  \begin{picture}(1,0.37107402)%
    \lineheight{1}%
    \setlength\tabcolsep{0pt}%
    \put(0,0){\includegraphics[width=\unitlength,page=1]{sketch_MonteCarlo_Method1.pdf}}%
    \put(0.02434005,0.09201147){\color[rgb]{0,0,0}\makebox(0,0)[lt]{\begin{minipage}{0.06810626\unitlength}\raggedright Laser\end{minipage}}}%
    \put(0.21409598,0.35767106){\color[rgb]{0,0,0}\makebox(0,0)[lt]{\begin{minipage}{0.55808506\unitlength}\raggedright Dispersive propagation medium\end{minipage}}}%
    \put(0.04190415,0.20699341){\color[rgb]{0,0,0}\makebox(0,0)[lt]{\begin{minipage}{0.1428714\unitlength}\raggedright Gaussian pulse\end{minipage}}}%
    \put(0,0){\includegraphics[width=\unitlength,page=2]{sketch_MonteCarlo_Method1.pdf}}%
    \put(0.25431351,0.10736838){\color[rgb]{0,0,0}\makebox(0,0)[lt]{\begin{minipage}{0.09364613\unitlength}\raggedright $L_0$\end{minipage}}}%
    \put(0,0){\includegraphics[width=\unitlength,page=3]{sketch_MonteCarlo_Method1.pdf}}%
    \put(0.35813955,0.04698368){\color[rgb]{0,0,0}\makebox(0,0)[lt]{\begin{minipage}{0.15057869\unitlength}\raggedright $(T_0, \Theta_0)$\end{minipage}}}%
    \put(0.50050729,0.15202034){\color[rgb]{0,0,0}\makebox(0,0)[lt]{\begin{minipage}{0.16175236\unitlength}\raggedright $(T_1, \Theta_1)$\end{minipage}}}%
    \put(0.6142869,0.33503451){\color[rgb]{0,0,0}\makebox(0,0)[lt]{\begin{minipage}{0.16175236\unitlength}\raggedright $(T_2, \Theta_2)$\end{minipage}}}%
    \put(0,0){\includegraphics[width=\unitlength,page=4]{sketch_MonteCarlo_Method1.pdf}}%
    \put(0.82436779,0.34037798){\color[rgb]{0,0,0}\makebox(0,0)[lt]{\begin{minipage}{0.11226893\unitlength}\raggedright Deformed pulse\\ \end{minipage}}}%
    \put(0.90680955,0.03294193){\color[rgb]{0,0,0}\makebox(0,0)[lt]{\begin{minipage}{0.09484563\unitlength}\raggedright Detector\\ \end{minipage}}}%
    \put(0,0){\includegraphics[width=\unitlength,page=5]{sketch_MonteCarlo_Method1.pdf}}%
    \put(0.71325995,0.16410712){\color[rgb]{0,0,0}\makebox(0,0)[lt]{\begin{minipage}{0.16175236\unitlength}\raggedright $(T_3, \Theta_3)$\end{minipage}}}%
    \put(0.87082832,0.25365166){\color[rgb]{0,0,0}\makebox(0,0)[lt]{\begin{minipage}{0.16175236\unitlength}\raggedright $T_g'$\end{minipage}}}%
    \put(0.43322304,0.15733482){\color[rgb]{0,0,0}\makebox(0,0)[lt]{\begin{minipage}{0.09364613\unitlength}\raggedright $L_1$\end{minipage}}}%
    \put(0.57348431,0.27697834){\color[rgb]{0,0,0}\makebox(0,0)[lt]{\begin{minipage}{0.09364613\unitlength}\raggedright $L_2$\end{minipage}}}%
    \put(0.7044149,0.26870111){\color[rgb]{0,0,0}\makebox(0,0)[lt]{\begin{minipage}{0.09364613\unitlength}\raggedright $L_3$\end{minipage}}}%
    \put(0.84813758,0.17539424){\color[rgb]{0,0,0}\makebox(0,0)[lt]{\begin{minipage}{0.09364613\unitlength}\raggedright $L_4$\end{minipage}}}%
    \put(0,0){\includegraphics[width=\unitlength,page=6]{sketch_MonteCarlo_Method1.pdf}}%
  \end{picture}%
\endgroup%

%% file: sketch_MonteCarlo_Method3.pdf_tex
\begingroup%
  \makeatletter%
  \providecommand\color[2][]{%
    \errmessage{(Inkscape) Color is used for the text in Inkscape, but the package 'color.sty' is not loaded}%
    \renewcommand\color[2][]{}%
  }%
  \providecommand\transparent[1]{%
    \errmessage{(Inkscape) Transparency is used (non-zero) for the text in Inkscape, but the package 'transparent.sty' is not loaded}%
    \renewcommand\transparent[1]{}%
  }%
  \providecommand\rotatebox[2]{#2}%
  \newcommand*\fsize{\dimexpr\f@size pt\relax}%
  \newcommand*\lineheight[1]{\fontsize{\fsize}{#1\fsize}\selectfont}%
  \ifx\svgwidth\undefined%
    \setlength{\unitlength}{796.79306681bp}%
    \ifx\svgscale\undefined%
      \relax%
    \else%
      \setlength{\unitlength}{\unitlength * \real{\svgscale}}%
    \fi%
  \else%
    \setlength{\unitlength}{\svgwidth}%
  \fi%
  \global\let\svgwidth\undefined%
  \global\let\svgscale\undefined%
  \makeatother%
  \begin{picture}(1,0.41133208)%
    \lineheight{1}%
    \setlength\tabcolsep{0pt}%
    \put(0,0){\includegraphics[width=\unitlength,page=1]{sketch_MonteCarlo_Method3.pdf}}%
    \put(0.02174789,0.11045464){\color[rgb]{0,0,0}\makebox(0,0)[lt]{\begin{minipage}{0.0608531\unitlength}\raggedright Laser\end{minipage}}}%
    \put(0.19129526,0.40089653){\color[rgb]{0,0,0}\makebox(0,0)[lt]{\begin{minipage}{0.3117977\unitlength}\raggedright Dispersive propagation medium\end{minipage}}}%
    \put(0.06691717,0.20843711){\color[rgb]{0,0,0}\makebox(0,0)[lt]{\begin{minipage}{0.12765592\unitlength}\raggedright Gaussian pulse\end{minipage}}}%
    \put(0,0){\includegraphics[width=\unitlength,page=2]{sketch_MonteCarlo_Method3.pdf}}%
    \put(0.22722971,0.12417607){\color[rgb]{0,0,0}\makebox(0,0)[lt]{\begin{minipage}{0.08367303\unitlength}\raggedright $L_0$\end{minipage}}}%
    \put(0,0){\includegraphics[width=\unitlength,page=3]{sketch_MonteCarlo_Method3.pdf}}%
    \put(0.34926331,0.06681274){\color[rgb]{0,0,0}\makebox(0,0)[lt]{\begin{minipage}{0.1345424\unitlength}\raggedright $\Theta_0$\end{minipage}}}%
    \put(0,0){\includegraphics[width=\unitlength,page=4]{sketch_MonteCarlo_Method3.pdf}}%
    \put(0.76562504,0.36590024){\color[rgb]{0,0,0}\makebox(0,0)[lt]{\begin{minipage}{0.10031255\unitlength}\raggedright Deformed pulse\\ \end{minipage}}}%
    \put(0.88745919,0.03132037){\color[rgb]{0,0,0}\makebox(0,0)[lt]{\begin{minipage}{0.08474479\unitlength}\raggedright Detector\\ \end{minipage}}}%
    \put(0.82425255,0.29506589){\color[rgb]{0,0,0}\makebox(0,0)[lt]{\begin{minipage}{0.14452611\unitlength}\raggedright $T_g'$\end{minipage}}}%
    \put(0.51336037,0.27762462){\color[rgb]{0,0,0}\makebox(0,0)[lt]{\begin{minipage}{0.08367303\unitlength}\raggedright $L_2$\end{minipage}}}%
    \put(0.63985547,0.25881896){\color[rgb]{0,0,0}\makebox(0,0)[lt]{\begin{minipage}{0.08367303\unitlength}\raggedright $L_3$\end{minipage}}}%
    \put(0,0){\includegraphics[width=\unitlength,page=5]{sketch_MonteCarlo_Method3.pdf}}%
    \put(0.4777998,0.16529636){\color[rgb]{0,0,0}\makebox(0,0)[lt]{\begin{minipage}{0.1345424\unitlength}\raggedright $\Theta_1$\end{minipage}}}%
    \put(0.66703472,0.1737965){\color[rgb]{0,0,0}\makebox(0,0)[lt]{\begin{minipage}{0.1345424\unitlength}\raggedright $\Theta_3$\end{minipage}}}%
    \put(0,0){\includegraphics[width=\unitlength,page=6]{sketch_MonteCarlo_Method3.pdf}}%
    \put(0.57874683,0.27083707){\color[rgb]{0,0,0}\makebox(0,0)[lt]{\begin{minipage}{0.1345424\unitlength}\raggedright $\Theta_2$\end{minipage}}}%
    \put(0,0){\includegraphics[width=\unitlength,page=7]{sketch_MonteCarlo_Method3.pdf}}%
    \put(2.9151769,0.15769828){\color[rgb]{0,0,0}\makebox(0,0)[lt]{\begin{minipage}{0.59295829\unitlength}\raggedright \end{minipage}}}%
    \put(0,0){\includegraphics[width=\unitlength,page=8]{sketch_MonteCarlo_Method3.pdf}}%
    \put(0.4070532,0.13934547){\color[rgb]{0,0,0}\makebox(0,0)[lt]{\begin{minipage}{0.08367303\unitlength}\raggedright $L_1$\end{minipage}}}%
    \put(0,0){\includegraphics[width=\unitlength,page=9]{sketch_MonteCarlo_Method3.pdf}}%
    \put(0.61404353,0.02191363){\color[rgb]{0,0,0}\makebox(0,0)[lt]{\begin{minipage}{0.1345424\unitlength}\raggedright Time\end{minipage}}}%
    \put(0.47681945,0.02016048){\color[rgb]{0,0,0}\rotatebox{90}{\makebox(0,0)[lt]{\begin{minipage}{0.1345424\unitlength}\raggedright Amplitude\end{minipage}}}}%
    \put(0,0){\includegraphics[width=\unitlength,page=10]{sketch_MonteCarlo_Method3.pdf}}%
    \put(0.80125159,0.2204835){\color[rgb]{0,0,0}\makebox(0,0)[lt]{\begin{minipage}{0.08367303\unitlength}\raggedright $L_4$\end{minipage}}}%
  \end{picture}%
\endgroup%